\documentclass[10pt]{amsart}
\usepackage[cp1251]{inputenc}
\usepackage[english,russian]{babel}
\usepackage{amsmath}
\usepackage{amssymb}
\usepackage{amsfonts}

\newcommand{\be}{\begin{equation}}
\newcommand{\ee}{\end{equation}}
\def\cN{{\mathcal N}}
\def\cD{{\mathcal D}}
\def\cM{{\mathcal M}}
\def\cA{{\mathcal A}}
\def\f{\frac}

\def\udcs{512.5} %Здесь автор определяет УДК своей работы
\def\mscs{13A99} %Здесь автор определяет классификаторы AMS своей работы

\def\logo{{\bf\huge S\raisebox{0.2ex}{\hspace{0.55ex}\raisebox{0.05ex}e\hspace{-1.65ex}$\bigcirc$}MR}}

\def\top
     {
  \vbox{
     \noindent\logo
     \hspace{80mm}\raisebox{1ex}{ISSN 1813-3304 }

     \vspace{5mm}

     \begin{center}
     {\huge СИБИРСКИЕ \ ЭЛЕКТРОННЫЕ} \\[2mm]
     {\huge МАТЕМАТИЧЕСКИЕ ИЗВЕСТИЯ} \\[2mm]
     {\large Siberian Electronic Mathematical Reports} \\[1mm]
     {\LARGE\tt{http://semr.math.nsc.ru}}\\[0.5mm]
     \end{center}
     \vspace{-3mm}
     \noindent
     \begin{tabular}{c}
     \hphantom{aaaaaaaaaaaaaaaaaaaaaaaaaaaaaaaaaaaaaaaaaaaaaaaaaaaaaaaaaaaaaaaaaaaaaa} \\
     \hline\hline
     \end{tabular}

     \vspace{1mm}
     {\flushleft\it Том 6, стр. 272--311 (2009) \hspace{65mm}{\rm\small УДК \udcs}}
     \newline
         \hphantom{aaaaaaaaaaaaaaaaaaaaaaaaaaaaaaaaaaaaaaaaaaaaaaaaaaaaaaaaaaaaaa}{\rm\small MSC\ \ \mscs }
  }%\vbox
}

\begin{document}

\thispagestyle{empty}

\title{Filippov--Nambu $n$-algebras  relevant to physics}
\author{{N.G. Pletnev}}%
\address{Николай Гаврилович Плетнев
\newline\hphantom{iii} Институт математики им. С.~Л.~Соболева СО РАН,
\newline\hphantom{iii} пр. академика Коптюга 4,
\newline\hphantom{iii} 630090, Новосибирск, Россия}%
\email{pletnev@math.nsc.ru}%

\thanks{\sc Pletnev, N.G.,
Filippov-Nambu $n$-algebras  relevant to physics}
\thanks{\copyright \ 2009 Плетнев Н.Г}
\thanks{\rm Работа поддержана грантами RFBR 09-02-00078, 08-02-00334-a  и NSh - 2553.2008.2.}
\thanks{\it Поступила July, 8, 2009, опубликована October, 16, 2009 }%

\top \vspace{1cm} \maketitle {\small
\begin{quote}
\noindent{\sc Abstract. } Gauge symmetry based on Lie algebra has a
rather long history and it  successfully describes electromagnetism,
weak and strong interactions in the nature. Recently the
Filippov--Nambu 3-algebras have been in the focus of interest since
they appear as gauge symmetries of new superconformal Chern--Simons
non-Abelian theories in 2 + 1 dimen\-sions with the maximum allowed
number of ${\cN} = 8$ linear super\-sym\-met\-ries. These theories
explore the low energy dynamics of the micro\-sco\-pic degrees of
freedom of coincident M2 branes and constitute the boundary
conformal field theories of the bulk $AdS_4 \times S_7$ exact
11-dimen\-sio\-nal supergravity backgrounds of supermembranes. These
mysterious new symmetries, the Filippov--Nambu 3-algebras represent
the imple\-men\-ta\-tion of non-asso\-cia\-tive algebras of
coordinates of charged tensionless strings, the boundaries of open
M2 branes in antisymmetric field magnetic backgrounds of M5 branes
in the M2 -M5 system. A crucial input into this construction came
from the study of the M2-M5 system in the Basu--Harvey’s work  where
an equation describing the Bogomol'nyi--Prasad--Sommerfield (BPS)
bound state of multiple M2-branes ending on an M5 was formulated.
The Filip\-pov--Nambu 3-algeb\-ras are either operator or matrix
representation of the classical Nambu symmetries of world volume
preserving diffeo\-morp\-hisms of M2 branes. Indeed at the classical
level the supermembrane Lagrangian, in the covariant formulation,
has the world volume preserving diffeo\-mor\-phisms symmetry
$SDiff(M_{2+1})$. The Filippov--Nambu 3-algeb\-ras presumably
correspond to the quantization of the rigid motions in this infinite
dimensional group, which describe the low energy excitation spectrum
of the M2 branes. It emphasizes the Filippov--Nambu n-algebras   as
the mathematical framework for describing symmetry properties of
classical and quantum mechanical systems.

\medskip

 \end{quote}
}

\section{Introduction}
Mathematics provides us with a language in which we formulate the
laws that govern  phenomena observed in the nature. This language
has proven to be both powerful and effective.

A foundation of physics cannot be built solely on this ground,
however; an even more essential ingredient is experiment, and any
substantial progress in physics  eventually has to lead to
predictions that can be tested experimentally. Neverthe\-less, the
quest for a deeper understanding of fundamental physical issues,
such as the interactions among elementary particles or the structure
of space-time, as results us to theories which are even harder to
put to observational tests. In this situation, mathematical
conciseness and internal consistency of a physical theory become
increasingly important guidelines in the evolution of physics. In
studies of physical phenomena we want to discover hidden
mathematical structures which govern underlying processes. These
structures can be either known or new but in any case new studies in
physics pose new mathematical problems even in old classical areas
of mathematics. In its turn, studies of mathematical structures
relevant to the physical phenomena lead to new developments of
physical theories. Textbook example providing an important role in
physics is an algebra and representation theory. They are the basis
of major progress in string theory, conformal and topolo\-gical
quantum field theory, and integrable systems. Conversely, ideas from
these areas are directly related to new developments in mathematics.

In recent years, novel issues such as aspects of three-dimensional
topological field theories which are expected to be relevant to
topological quantum computing and string/${\cM}$ theory have merged
in mathematical physics, especially in quantum field theory.
Accordingly, additional areas of mathematics have become influential
and, in turn, been influenced themselves by the developments in
physics.

Any search for generalizations of synthesis of quantum mechanics and
relativity theory has to have a well defined motivation. One
possible general starting point is provided by the observation that
the evolution of fundamental physical theories, characterized by
appearance of new dimensionful parameters (new constants of nature),
can be mathematically understood from the point of view of
deformation theory. In particular, relativity theory, quantum
mechanics and quantum field theory can be understood mathematically
as deformations of unstable structures \cite{deform}\footnote{An
algebraic structure is called stable (or rigid) for a class of
deformations if any deformation in this class leads to an equivalent
(isomorphic) structure.}. An example of an unstable algebraic
structure is non-relativistic classical mechanics. By deforming an
unstable structure, such as classical non-relativistic mechanics,
via dimensionful deformation parameters, the speed of light $c$ and
the Planck constant $\hbar$, one obtains new stable structures -
special relativity and quantum mechanics. Likewise, relativistic
quantum mechanics (quantum field theory) can be obtained through a
double ($c$ and $\hbar$) deformation. It is natural to expect that
there is a further deformation via one more dimensionful constant,
the Planck length  $l_P$. The resulting structure could be expected
to form a stable structural basis for a quantum theory of gravity. A
closely related idea has appeared in open string field theory, as
originally formulated by Witten \cite{witten86}. There, the
deformation parameters are the tension in the string $\alpha'$ and
$\hbar$. The classical open string field theory Lagrangian is based
on the use of the string field (which involves an expansion to all
orders in $\alpha'$) and a  star product which is defined in terms
of the world-sheet path integral, also involving $\alpha'$. The full
quantum string field theory is thus, in principle, an example of a
one-parameter ($\alpha'$) deformation of quantum mechanics. String
theory is well known to be the leading prospect for quantizing
gravity and unifying it with other interactions. One may also take a
broader view of string theory as a description of string-like
excitations that arise in many different physical systems, such as
the superconducting flux tubes or the chromo-electric flux tubes in
non- Abelian gauge theories. From the point of view of quantum field
theories describing the physical systems where these string-like
objects arise, they are ”emergent” rather than fundamental.

In string theory \cite{gsw} the graviton and all other elementary
particles are one-dimensional objects: strings, rather than points
as in quantum field theory. String theory may resemble the real
world in its broad outlines, but a decisive test still seems to be
far away. The main problem is that while there is a unique theory,
it has an enormous number of classical solutions, even if we
restrict attention to the solution with four large flat dimensions.
Upon quantization, each of these solutions is a possible ground
state for the theory and the four-dimensional physics is different
in each of these. Until recently, our understanding of different
versions of string theory was limited to perturbation theory, that
corresponds to small numbers of strings interacting weakly. It was
not known even how define the theory at strong coupling. String
theory was revolutionized by the discovery of $D$-branes
\cite{polch}. The understanding of these nonperturbative objects
allowed to uncover a deep connection between non-Abelian gauge
theories and string theory. This resulted in the Maldacena
correspondence \cite{maldac} where a fascinating duality between
gauge theories and string/gravitational theories is of great
importance. The origin of the non-Abelian degrees of freedom came
from the open strings extending between different $D$-branes and
becoming massless when the $D$-branes coincide. In addition, thanks
to the AdS/CFT correspondence and its extensions, we now know that
at least some field theories have dual formulations in terms of
string theories in curved backgrounds. In these examples, the
strings that are ''emergent'' from the field theory point of view
are dual to fundamental or $D$-strings in the string theoretic
approach. Besides being of great theoretical interest, such
dualities are becoming a useful tool for studying strongly coupled
gauge theories. These ideas also have far-reaching implications for
building connections between string theory and the real world.

In the past few years  a great  progress on these issues, we
observed connected largely with the systematic application of the
constraints imposed by supersymmetry. It was found that the strongly
coupled limit of any string theory is described by a dual weakly
coupled string theory, or by a new eleven-dimensional theory known
as $\cM$-theory whose low energy limit is an eleven dimensional
supergravity. The extended objects are no longer strings but
membranes and five-branes. All the different string theories are
different compactification limits of this single theory, as such
$\cM$-theory unified string theories. The five different versions of
string theory are just $\cM$-theory expanded around different vacua.
This $\cM$-theory web then explained the nonperturbative dualities
that had been conjectured in string theory some years before.
Understanding the structure of $\cM$-theory as an underlying theory
of all known string theories was one of the major efforts during the
past decade. Although several approaches have been found,  we  still
don't have any clear picture of this theory. A true formulation of
$\cM$-theory away from the low energy limit is still a far away
dream. Fortunately, in the last few years, some ground breaking
ideas were invented for describing  the dynamics of fundamental
objects of $\cM$-theory based on the Filippov 3-algebras \cite{fil},
i.e. membranes and five branes. The original motivation was a search
for a theory describing coincident M2-branes \cite{basu} (see also
\cite{berman} for the references).

While string  theories are based on two-algebra structure, recent
advances in  M2-theory started by Bagger--Lambert--Gustavsson (BLG)
\cite{BLG} suggest that full description of $\cM$-theory may require
a generalized Lie-algebra structure: namely three algebra or even
higher, $n$-algebra structure. In fact, the digits, two and three,
appear to have intriguing associations with string and $\cM$-theory
respectively, first of all, two is the dimension of string
worldsheet while three is that of membrane worldvolume. In the
mathematical literature, the Lie 3-algebra (a term originally coined
by Filippov \cite{fil}, following  earlier work of Nambu
\cite{Nambu}) is not new, and its structure has been studied to some
extent. In the long history of the study of the Nambu brackets their
relation to the supermembranes or $\cM$-theory  especially
interestingly. There have been many attempts to quantize the
classical the Nambu bracket towards this direction. However, since
the quantization is difficult and does not seem to be unique, we
need to understand which properties are essential from the physical
viewpoint. Recently there was some progress in constructing a $2+1$-
dimensional local quantum field theory with $SO(8)$ superconformal
symmetry \cite{BLG}. This is a useful significant step to obtain a
world-volume Lagrangian description for coincident M2-branes.
Crucial for the construction is the use of 3-algebras which are
built around antisymmetrized product of three operators. In general,
$n$-algebras as a natural generalization of Lie algebras are defined
by a multi-linear map $[\star, \ldots, \star]$: ${\cA}^n\rightarrow
{\cA}$ on a linear space ${\cA}=\sum v_a T_a, v_a \in \mathbb{C}.$

More than three decades ago, Nambu \cite{Nambu} proposed a
generalization of classical Hamiltonian mechanics. In his formalism,
he replaced the usual pair of canonical variables of the Hamiltonian
mechanics by a triplet of coordinates in an odd dimen\-sio\-nal
phase space. Furthermore, he formulated his dynamics using a ternary
opera\-tion, the Nambu bracket, as opposed to the usual binary
Poisson bracket. Yet the fundamental principles of a canonical form
of the Nambu's mechanics, similar to the invariant geometrical of
the Hamiltonian mechanics, have only recently been discovered
\cite{takh}. The re-emergence of this little known theory is
possibly due to its relevance to the recent mathematical structures
having their basis in the classical motion of topological open
membranes as well as maximally superintegrable systems, such as the
Hydrogen atom and so on that are controlled by classical Nambu
brackets. Since the basic idea of the Nambu mechanics is to extend
the usual binary operation on the phase space to multiple operations
of higher order, this theory may also give some insights into the
theory of higher order algebraic structures and their possible
physical significance. In any case, both the Nambu's and Filippov's
works motivated and inspired a lengthly survey of these ideas by
Takhtajan \cite{takh} (also see \cite{vais99} ) along with many
other studies.

In these review we have described main features and problems of
 a consistent $\cM$-theory and pointed out some ways using
the concepts of $n$-algebras along which it has been developed over
last years. This survey is intended for  mathematicians who are
non-specialists in the field of theoretical physics. Therefore,
technical details are kept to a minimum and we refer to various
other literature throughout for the relevant formalisms. There are
many other topics in this vast field that are not touched upon in
this review. Furthermore, the bibliography is not meant to be
exhaustive, and we apologise  in advance to those concerned for the
omissions. For more details related to mathematical aspects of the
 properties and applications of certain n-ary
generalizations of Lie algebras in a self-contained and unified way
we refer to reviews \cite{AZC10}.

\section{$n$-Lie algebras}
We will briefly recall the definition of metric $n$-Lie algebras as
introduced by Filippov in \cite{fil}, of which ordinary Lie algebras
(n=2) and the 3-algebras appearing in the BLG theory and the Nambu
3-algebras \cite{Nambu} are special cases.

Define a (complex) $n$-Lie algebra as an algebra with an $n$-ary map
$[\cdot, \ldots, \cdot]$: ${\cA}^n \rightarrow {\cA}$ such that:

(a) $[\cdot, \ldots, \cdot]$ is totally antisymmetric, i.e. \be
[T_1, \ldots,T_n]=(-1)^{\varepsilon(\sigma)}[T_{\sigma(1)}, \ldots,
T_{\sigma(n)}], \quad T_i \in {\cA},\ee for all $T_1,\ldots, T_n \in
\cA$ and $\varepsilon(\sigma)$ is the parity of a permutation
$\sigma$.

(b) any $(n-1)$-plet acts via $[\cdot, \ldots, \cdot]$ as a
derivative, i.e. the bracket satisfies the fundamental identity for
all $T_i, R_i \in {\cA}$ \be [T_1, \ldots, T_{n-1},[R_1,
\ldots,R_n]]=\sum_{i=1}^n[R_1,\ldots,R_{i-1},[T_1,\ldots, T_{n-1},
R_i], R_{i+1}, \ldots, R_n] ,\ee which preserves main properties of
the Jacobi identity. It means that the bracket $[T_1, \ldots,
T_{n-1}]$ acts as a derivative on $\cA$, and it may be represented a
symmetry transformation. In terms of the basis, $n$-algebra is
expressed in terms of  structure constants \be\label{struct_const}
[T_{a_1}, \ldots, T_{a_n}]=if_{a_1\ldots a_n}^{\quad\quad b} T_b.\ee
The fundamental identity implies a bilinear relation between the
structure constants \be\label{FI}\sum_c f_{b_1 \ldots\,
b_p}^{\quad\quad c}f_{a_1\ldots \,a_{p-1}c}^{\quad\quad \quad d}=
\sum_i\sum_c f_{a_1 \ldots\, a_{p-1} b_i}^{\quad \quad \quad
c}f_{b_1\ldots\, c\ldots\, b_p}^{\quad\quad\quad d} .\ee The adjoint
action of $\wedge^{n-1}{\cA} $ on $\cA$ is defined as follows
$$ad_\Lambda v=f^{a_1\ldots a_p}_{\ \ \ \ \ \ b}\Lambda_{a_1\ldots  \;a_{p-1}}v_p T^b,$$
where $v=v_aT^a$ and $\Lambda=\Lambda_{a_1\ldots
\;a_{p-1}}T^{a_1}\wedge \ldots \wedge T^{a_{p-1}}.$ The fundamental
identity is equivalent to the statement that the adjoint action acts
as a derivative on the bracket
$$ad_\Lambda([v_1, \ldots, v_p])=[(ad_\Lambda v_1), \ldots,
v_p]+\ldots +[v_1, \ldots, (ad_\Lambda v_p)].$$ The derivatives
$ad_\Lambda$ obviously form a Lie algebra $ad_\Lambda
ad_{\tilde\Lambda}-ad_{\tilde\Lambda}
ad_\Lambda=ad_{[\Lambda,\tilde\Lambda]}.$

The $n$-Lie algebra can be equipped with an invariant inner product
as a bilinear map from $\cA \times \cA$ to $C$ \be\label{metr}<T_a,
T_b>=h_{ab}.\ee We will refer to the symmetric tensor $h_{ab}$ as
the metric. As a generalization of the Killing form in Lie algebra,
we require that the metric is invariant under any transformation
generated by the bracket $[T_{a_1}, \ldots, T_{a_{n-1}}, \cdot]$:
\be<[T_{a_1}, \ldots, T_{a_{n-1}}, T_b], T_c> + <T_b, [T_{a_1},
\ldots, T_{a_{n-1}}, T_c]>=0~.\ee This implies a relation for the
structure constants \be h_{cd}f_{a_1\ldots\,a_{n-1}
b}^{\quad\quad\quad d}+h_{bd}f_{a_1\ldots\,a_{n-1}
c}^{\quad\quad\quad d}=0~,\ee therefore the tensor \be\label{anti}
f_{a_1\ldots a_n}\equiv f_{a_1\ldots\,a_{n-1} }^{\quad\quad\quad
b}h_{ba_n},\ee is totally antisymmetrized. For applications to
physics, it is very important to have a nontrivial metric $h_{ab}$
in order to write down the Lagrangian or physical observables which
are invariant under transformations defined by $n$-brackets.
Assuming the positivity of metric $h$ leads to severe restrictions
on the structure constants of $n$-algebra \cite{no-go}. The adjoint
action of a $n$-algebra with an invariant metric can be described
alternatively through the matrix action on $\cA.$ The element
$\Lambda\in \wedge^{n-1}{\cA}$ can be mapped to $Mat_{n\times n},$
$n=\mbox{dim}{\cA}$ as follows $\lambda^c_{\ \ b}=f^{a_1\ldots
a_{n-1}c}_{\ \ \ \ \ \ \ \ \  \ \ b}\Lambda_{a_1\ldots a_{n-1}},$
such that $\lambda$'s satisfy the following properties
$$f^{a_1\ldots a_n}_{\ \ \ \ \ \ c}\lambda^c_{\ \ b}=
f^{c a_2\ldots a_n}_{\ \ \ \ \ \ b}\lambda^{a_1}_{\ \ c}+\ldots
+f^{a_1\ldots a_{n-1}c}_{\ \ \ \ \ \ c} \lambda^{a_n}_{\ \ c}, \quad
\lambda^a_{\ \ c}h^{cb}=-h^{ac}\lambda^b_{\ \ a}.$$

Another mathematical structure of physical importance is the
Hermitian conjuga\-tion. A natural definition of the Hermitian
conjugate of an $n$-bracket is \be\label{herm}[A_1, \ldots
,A_n]^\dag = [A^\dag_ n, \ldots ,A^\dag_ 1].\ee This relation
determines the reality of the structure constants.  If we choose the
generators to be Hermitian for the usual Lie algebra,, the structure
constants $f_{ab}^ {\;\;c}$ are real numbers, and if the generators
are anti-Hermitian, the structure constants are imaginary. This is
not the case for 3-brackets. The structure constants are always
imaginary when the generators are all Hermitian or all
anti-Hermitian. In general, for $n$-brackets, the structure
constants are real if $n = 0, 1$ (mod 4).They are  imaginary if $n =
2, 3$ (mod 4) for the Hermitian generators. The structure constants
are multiplied by a factor of $\pm i$ when we replace the Hermitian
generators by anti-Hermitian ones only for even $n$.

Simple examples are given by the $n$-Lie algebras \cite{fil}. In
particular, it is shown that vector multiplication of vectors of the
$(n+1)$-dimensional Euclidean space and the Jacobian $|\partial
f_i/\partial x_j|$ of the polynomials $f_1,\ldots, f_n \in {\cA}$
algebra in $n$ variables $x_1, \ldots, x_n$ on an oriented
$n$-dimensional manifold can be taken as canonical examples of the
operations in ${\cA}$. However, it is not only that the complete
classification of the $n$-algebra does not exist, but there are very
few explicit examples in the literature.

\subsection{Three algebras and $\cN=8,6$ Chern--Simons gauge theories}

The branes of $\cM$-theory are important but still  are quite
mysterious objects. Recently the construc\-tion of superconformal
Chern--Simons--matter theories in three dimensions has attrac\-ted a
lot  of attention in string/$\cM$-theory community, because they are
natural candi\-da\-tes for the dual gauge description of M2 branes
in $\cM$ theory \cite{schwarz04}. Briefly, there are eleven bosonic
degrees of freedom corresponding to the embedding of the membrane.
The reparameterization invariance of the worldvolume gauges away
three of these so that there are eight bosonic degrees of freedom at
the end. The fermions start out as thirty-two component spinors. The
mass-shell condition and the $\kappa$-symmetry each have the
available degrees of freedom. The bosonic and fermionic degrees of
freedom are then organized as an ${\cN} = 8$ multiplet of the
three-dimensional worldvolume theory.

As is well known,  generically Chern--Simons gauge theories in three
dimensions are conformally invariant, both for pure gauge theories
and for theories coupled to massless matter fields. This remains
true even at the quantum level (in spite of a quantum shift at one
loop order), the Chern--Simons gauge coupling does not run at all,
because its $\beta$ function vanishes, as shown both by an explicit
two-loop calculations for theories with matter and by formal proof
up to all orders in perturbation theory for pure gauge theories. In
order to construct the dual gauge description of M2 branes, the
relevant issue is then how to incorporate extended supersymmetries
into Chern--Simons-matter theories, since extended supersymmetry
plays a crucial role in $\cM$-theory as it does in superstring
theory.

In a series of recent papers \cite{BLG} a non-Abelian model of
multiple M2-branes based on an 3-algebra as the internal symmetry
has been proposed. The theory living on an M2 brane is conformal. So
the fields acquire the length dimensions \be\{A_m, X^a, \Psi,
\epsilon\}=\{-1, -\f12, -1,\f12\}.\ee One may ask what requirements
come from supersymmetry. The fermionic field $\Psi$ is a Majorana
spinor in $10+1$ dimensions satisfying the chirality condition
$\Gamma_{012}\Psi=-\Psi$. As result $\Psi$ has 16 real fermionic
components, equivalent to 8 bosonic degrees of freedom $X^I$. In
$2+1$ dimensions  a gauge potential usually has one propagating
degree of freedom. However, here the gauge potential has no
canonical kinetic term, but only the Chern--Simons term, and hence
it has no propagating degrees of freedom. Simple dimensional
analysis suggests that in the supersymmetry variations include
product of two as well as of three fields. It is of course desirable
that all products of our fields are such that they close on some
internal algebra. The way  do that is to making the minimal
assumption that there is a multiplication of two and three fields
which belong to some set of fields, that we denote as $\cA$, such
that the product of three elements in $\cA$ must yield back an
element in $\cA$.  Then we see what requirements of closure of the
supersymmetry transformations impose these new type multiplications.

Based on the totally antisymmetric Filippov 3-brackets, the
maximally (i.e. $\cN$=8) supersymmetric Chern-Simons-matter theory
in d3 with $SO(4)$ gauge group and $SO(8)$ R-symmetry, was
constructed \cite{BLG} as the dual gauge description of two M2
branes. It was also shown that in possible to overcome the obstacles
of no-go theorem \cite{no-go}  only for three-algebra which has a
symmetric and positive defined metric is either $so(4)$ or direct
sum of a number of $so(4)$'s. In this sense the BLG theory is rather
unique and can describe only two coincident M2-branes. The BLG
theory is based on  3-algebras. A 3-algebra $\cA$ is an $N$
dimensional vector space endowed with a trilinear skew-symmetric
product $[A,B,C],$ which satisfies the so called fundamental
identity \be\label{fi3} [A,B,[C,D,E]]=[[A,B,C],
D,E]+[C,[A,B,D],E]+[C,D,[A,B,E]]~. \ee If we let $\{T^a\}_{1\leq
a\leq N}$ to be a basis of $\cA$, then the 3-algebra will be
specified by the structure constants $f^{abc}_{\ \ \ d}$ of $\cA$:
\be [T^a,T^b,T^c]=f^{abc}_{\ \ \ d} T^d.\ee The fundamental identity
(\ref{fi3}) is expressed as: \be\label{fi-3f} f^{abg}_{\ \ \
h}f^{cde}_{\ \ \ g}=f^{abc}_{\ \ \ g}f^{gde}_{\ \ \ h}+f^{abd}_{\ \
\ g}f^{cge}_{\ \ \ h}+f^{abe}_{\ \ \ g}f^{cdg}_{\ \ \ h}.\ee
Classifying 3-algebra $\cA$ requires classifying the solutions of
the fundamental identity (\ref{fi-3f}) for the structure constants
$f^{abc}_{\ \ \ d}.$ In order to derive from the Lagrangian
description the equations of motion of the BLG theory a bi-invariant
non-degenerate metric  (\ref{metr}) that arises by postulating a
bilinear scalar product $\mbox{Tr}(.,.)$ on the 3-algebra is needed
\be h^{ab}=\mbox{Tr}(T^a,T^b).\ee The Lagrangian of the BLG theory
is completely specified once a collection of structure constants
$f^{abc}_{\ \ \ d}$ and a bi-invariant metric $h^{ab}$
are given. The BLG theory encodes the interactions of a three dimensional $\cN=8$ multiplet, consisting of eight %($I=1, \ldots 8$)
scalar fields $X^I$ and their fermionic superparthers $\Psi$, and a
non-propagating gauge field $A_{m b}^{\ \ a}.$ Matter fields in this
theory take values in $\cA$, so that $X^I=X^I_a T^a,$
$\Psi=\Psi_aT^a.$ The indices $I,J,K$ run in $1, \ldots,8,$ and they
specify the transverse directions of M2-brane; we denote the
world-volume of the membrane as $\cM$ and its longitudial directions
as $x^m$ where $m,n$ run in $0,1,2$. The indices $a,b,c$ take values
in $1,\ldots, {N}$ where ${N}$ is the number of generators of the
Lie 3-algebra specified by a set of structure constants $f^{abc}_{\
\ \ d}.$ The fermionic field $\Psi$ is a Majorana spinor in $10+1$
dimensions and $\Gamma^M =\{\gamma^m, \Gamma^I\}$ are
eleven-dimensional gamma matrices satisfying the Clifford algebra
$\{\Gamma^M,\Gamma^N\}=2\eta^{MN}.$ As result $\Psi$ has 16 real
fermionic component, equivalent to 8 bosonic degrees of freedom.

The BLG Lagrangian is given by \cite{BLG} \be{\mathcal
L}=-\f12\cD_mX_{aI}\cD^m X^I_a+\f{i}{2}\bar\Psi^a\gamma^m\cD_m\Psi_a
+\f{i}{4}f_{abcd}\bar\Psi^b\Gamma^{IJ}X^{cI}X^{dJ}\Psi^a\ee
$$-\f{1}{12}(f_{abcd}X^{aI}X^{bJ}X^{cK})(f_{efg}^{\ \ \ d}X^{eI}X^{fJ}X^{gK})+\f12\varepsilon^{mnl}(f_{abcd}A^{ab}_m\partial_n A_l^{cd})$$$$+
\f23f_{aef}^{\ \ \ g}f_{bcdg}A^{ab}_mA^{cd}_nA^{ef}_l,$$ where: \be
\cD_m X^{aI}=\partial_mX^{aI}+f^a_{\ bcd}A^{cd}_mX^{bI}.\ee The
theory is invariant under the gauge transformations \be\delta
X^{aI}=-f^a_{\ \ bcd}\Lambda^{bc}X^{dI},\ee
$$\delta \Psi^a=-f^{a}_{\ \ bcd}\Lambda^{bc}\Psi^d,$$
$$\delta (f_{ab}^{\ \  cd}A^{ab}_m)=f_{ab}^{\ \  cd}\cD_m\Lambda^{ab},$$
and under the following supersymmetry transformations
\be\label{susytrans}\delta X^{aI}=i\bar\epsilon\Gamma^I\Psi^a,\ee
$$\delta\Psi^a=\cD_mX^{aI}\gamma^m\Gamma^I\epsilon +\f16f^{a}_{\ \ bcd}X^{bI}X^{cJ}X^{dK}\Gamma^{IJK}\epsilon,$$
$$\delta (f_{ab}^{\ \  cd}A^{ab}_m)=if_{ab}^{\ \ cd}X^{aI}\bar\epsilon\gamma_m\Gamma_I\Psi^b,$$
where $\Psi$ and $\epsilon$ are  16-component Majorana spinors
satisfying the projection condi\-tion
$\gamma_{012}\epsilon=\epsilon$ and $\gamma_{012}\Psi^a=-\Psi^a$
respectively.

When $h^{ab}$ is positive definite the only one known example of
this algebraic structure was given in \cite{BLG}. In this case, the
vector space is $\mathbb{R}^4$ and we can take \be
h^{ab}=\delta^{ab}, \quad f^{abcd}=f \epsilon^{abcd},\ee for some
constant $f$. The usual constraint that arises by demanding
invariance under large gauge transformations  requires us to choose
$f=\f{2\pi}{\kappa}$ where the level $\kappa$ is an integer. Then
the triple product is the natural generalization to four dimensions
of the usual cross product: it gives a new vector perpendicular to
the vectors in the product whose length is the signed of the
parallelepiped spanned by the vectors.

Recently, there have been several attempts to relax these
assumptions and const\-ruct additional field-theory models of
multiple M2-branes. There have been interes\-ting proposals in which
the metric $h^{ab}$ has a Lorentzian (indefinite) signature
\cite{Lor}. This allows one to construct an associated 3-algebra for
any Lie algebra, and the corresponding $\cN=8$ superconformal at the
classical level Lagrangian. Although these models are built on a
3-algebra without a positive norm and have pathologic ghost-type
fields (fields with negative kinetic energy), the corresponding
quantum theories have been argued to be unitary and they have some
encouraging features \cite{quantLor}. Choosing the Lorentzian
metric, one finds an infinite class of 3-algebras ${\cA}_{\mathcal
G}$ with an underlying Lie algebra structure \cite{quantLor}. For
any Lie algebra $\mathcal G$, $[T^i,T^j]=f^{ij}_{\ \ k}T^k$ with
structure constants $f^{ij}_{\ \ k}$ and Killing form $h^{ij}$ one
can definite the corresponding 3-algebra as follows. Let the
generators $T^a$ of the 3-algebra be denoted by $T^-,T^+,T^i$
$(a=+,-,i;i=1,\ldots, \mbox{dim}{\mathcal G})$, where $T^i$ are in
one-to-one correspondence with the generators of the Lie algebra.
Then the basic 3-algebra relations are chosen to be
$$[T^-,T^a;T^b]=0, \quad [T^+,T^i;T^j]=f^{ij}_{\ \ k}T^k,\quad  [T^i,T^j;T^k]=-f^{ijk}T^-~.$$
This set of totally antisymmetric structure constants  solves the
fundamental identity, where $f^{ijk}$ are structure constants of a
compact semi-simple Lie algebra $\mathcal G$ of dimension $n$. The
invariant inner product is defined as follows
$$<T^\mp,T^\mp>=0, \quad <T^-,T^+>=1,\quad <T^\mp,T^i>=0, \quad <T^i,T^j>=h^{ij}.$$
The Lagrangian based on ${\cA}_{\mathcal G}$, is gauge and
supersymmetry transformations free out the gauge fields $A_m^{+-}$
and $A^{+b}_m$, therefore they are not part of the theory.
Similarly, $A_{m ij}$ appears only through the combination
$f^{ijk}A_{m jk}=B^i_m$, so $A^i_m=A^{-i}_m, B^i_m$ will be viewed
as the fundamental gauge fields in the theory. The BLG Chern--Simons
term reduces, in this case of 3-algebra with Lorentzian signature to
a three dimensional $BF$ term. The peculiar form of the interactions
makes this model resemble, in some aspects the Yang--Mills theories
based on non semi-simple gauge groups \cite{Ts95}. In particular,
the quantum effective action contains only 1-loop term with the
divergent part that can be eliminated by a field redefinition. The
on-shell scattering amplitudes are thus finite (scale invariant).
This is a consequence of the presence of a null direction in the
field space metric: one of the field components is a Lagrange
multiplier which ‘freezes out’ quantum fluctuations of the
‘conjugate’ field. The non-positivity of the metric implies that
these theories are apparently non-unitary. However, the special
structure of interaction terms (degenerate compared to non-compact
Yang--Mills theories) suggests that there may exist a unitary
‘truncation’.

Another option is to look for theories with a reduced number of
supersymmetries. In \cite{Gai} a class of Chern--Simons Lagrangians
with $\cN=4$ supersymmetry was constructed. Of special interest is
the work \cite{ABJM} in which an infinite class of brane
configurations on the $\mathbb{C}^4/\mathbb{Z}_\kappa$ orbifold was
given whose low energy effective Lagrangian is the Chern--Simons
superconformal theory with $SO(6)$ R-symmetry and $\cN=6$
supersymmetry was constructed. The field content of the ABJM model
is given by four complex scalar and spinor fields which live in the
bifundamental representation of the $U(N)\times U(N)$ gauge group
while the gauge fields are governed by Chern--Simons actions of
levels $\kappa$ and $-\kappa$, respectively. Many aspects of the
$\cN=6$ theory have been studied \cite{Hai}, adding another evidence
for the existence of the M5-branes in the $\cN=6$ theory.

Thus it is of interest to generalize the construction based on
3-algebras on a complex vector space to the case of $\cN=6$
supersymmetry \cite{BL-N6}. This can be accomplished by relaxing the
conditions on the triple product
\be[T^a,T^b;\bar{T}^{\bar{c}}]=f^{ab\bar{c}}_{\ \ \ d}T^d,\ee so
that it is no longer real and antisymmetric in all three indices.
Rather it is required to satisfy \be
f^{ab\bar{c}\bar{d}}=-f^{ba\bar{c}\bar{d}}, \quad
f^{ab\bar{c}\bar{d}}=f^{\star\bar{c}\bar{d}ab}.\ee The triple
product is also required to satisfy the fundamental identity
\be\label{fi-N6} f^{ef\bar{g}}_{\ \ \ b}f^{cb\bar{a}}_{\ \ \
d}+f^{fe\bar{a}}_{\ \ \ b}f^{cb\bar{g}}_{\ \ \ d} +f^{\star
\bar{g}\bar{a}f}_{\ \ \ \ \ \bar{b}}f^{ce\bar{b}}_{\ \ \ d} +
f^{\star \bar{a}\bar{g}e}_{\ \ \ \ \ \bar{b}}f^{cf\bar{b}}_{\ \ \
d}=0.\ee To construct a gauge invariant Lagrangian   it is necessary
to have an inner product \be
h^{\bar{a}b}=\mbox{Tr}(\bar{T}^{\bar{a}}, T^b).\ee Then further
restrictions of $\cN=6$ supersymmetry, scale invariance, $SU(4)$
R-symmetry, and a global $U(1)$ give the conditions on the structure
constants $f^{ab\bar{c}\bar{d}}$. We use complex notation in which
the   supercharges $\epsilon_{AB}$ ($A=1, \ldots, 4$) are in the
representation $\bf 6$ of the algebra $SU(4)$ with vanishing $U(1)$
charge. They satisfy the reality condition
$\epsilon^{AB}=\f12\varepsilon^{ABCD}\epsilon_{CD}$. We introduce
four complex 3-algebra valued scalar fields $Z^A_a$ as well as their
complex conjugates $\bar{Z}_{A\bar{a}}$. Similarly, we denote the
fermions by $\Psi_{Aa}$ and their complex conjugates by
$\Psi_{\bar{a}}^A$. A raised $A$ index indicates that the field is
in the $\bf 4$ of $SU(4)$; a lowered index transforms in the
$\bar{\bf 4}$. We assign $Z^A_a$ and $\Psi_{Aa}$ a $U(1)$ charge of
1. Complex conjugation raises or lowers the $A$ index, flips the
sign of the $U(1)$ charge, and interchanges $a \leftrightarrow
\bar{a}$. We postulate the following supersymmetry transformations
\be\label{N6susy} \delta Z_a^A=i\bar\epsilon^{AB}\Psi_{Ba},\ee
$$\delta\Psi_{Bd}=\gamma^m\cD_mZ^A_d\epsilon_{AB}+f^{a\bar{b}c}_{1\ \  d}Z^C_a\bar{Z}_{C\bar{b}}Z^A_c\epsilon_{AB}
+f^{ab\bar{c}}_{2\ \ d}Z^C_aZ^D_b\bar{Z}_{B\bar{c}}\epsilon_{CD},$$
$$\delta \tilde{A}^{\ \ c}_{m \ d}=i\bar\epsilon_{AB}\gamma_mZ^A_a\Psi^B_{\bar{b}}f^{a\bar{b}c}_{3 \ \ d}
+i\bar\epsilon^{AB}\gamma_m\bar{Z}_{A\bar{a}}\Psi_{Bb}f^{\bar{a}bc}_{4
\ \ d},$$ where $f^{a\bar{b}c}_{1\ \  d}$, $f^{ab\bar{c}}_{2\ \ d}$,
$f^{a\bar{b}c}_{3 \ \ d}$ and $f^{\bar{a}bc}_{4 \ \ d}$ are tensors
of the 3-algebra. The covariant deriva\-ti\-ve is defined by $\cD_m
Z^A_d=\partial_mZ^A_d-\tilde{A}^{\ \ c}_{m \ \ d}Z^A_c$. Next, we
consider the closure of (\ref{N6susy}) on the scalars. Then we find
that $[\delta_1,\delta_2]Z_d^A$ only closes on to translations and a
gauge symmetry if $f^{a\bar{b}c}_{1 \ \ \ d}=f^{ac\bar{b}}_{2 \ \ \
d}$. Next, we examine the closure of the algebra on the fermions,
that gives $f^{\bar{a}bc}_{4 \ \ \ d}=-f^{b\bar{a}c}_{3 \ \ \ d}$
and $f^{a\bar{b}c}_{3 \ \ \ d}=f^{ac\bar{b}}_{2 \ \ \ d}$. Finally
the closure of (\ref{N6susy}) onto translations and gauge
transformations for the gauge field is succeed when
$f^{ab\bar{c}}_{2\ \ \ d}$ satisfies the fundamental identity
(\ref{fi-N6}) and that $\cD_m(f^{ab\bar{c}}_{2\ \ \ d})=0$. This is
just the statement that $f^{ab\bar{c}}_{2\ \ \ d}$ is an invariant
tensor of the gauge algebra.

With these results, it is not difficult to show that an invariant
Lagrangian is of the Chern--Simons form with interacting scalars,
fermions  and vectors that take values in a  3-algebra. As with the
$\cN=8$ model, the Lagrangian is entirely determined by specifying
of a triple product on a 3-algebra that satisfies the fundamental
identity and can be written as: \be\label{Lagr6} {\mathcal
L}=-\mbox{Tr}(\cD^m\bar{Z}_A,
\cD_mZ^A)-i\mbox{Tr}(\bar\Psi^A,\gamma^m\cD_m\Psi_A)-V+{\mathcal
L}_{CS}\ee
$$-i\mbox{Tr}(\bar\Psi^A,[\Psi_A,Z^B;\bar{Z}_B])+2i\mbox{Tr}(\bar\Psi^A,[\Psi_B, Z^B;\bar{Z}_A])$$
$$+\f{i}{2}\varepsilon_{ABCD}\mbox{Tr}(\bar\Psi^A,[Z^C,Z^D;\Psi^B])-\f{i}{2}\varepsilon^{ABCD}\mbox{Tr}(\bar{Z}_D,[\bar\Psi_A,\Psi_B;\bar{Z}_C]),$$
where the scalar potential \be V=\f23\mbox{Tr}(\Upsilon^{CD}_B,
\bar\Upsilon^B_{CD}),\ee
$$
\Upsilon^{CD}_B=[Z^C,Z^D;\bar{Z}_B]-\f12\delta^C_B[Z^E,Z^D;\bar{Z}_E]+\f12\delta_B^D[Z^E,Z^C;\bar{Z}_E],$$
and ${\mathcal L}_{CS}$ is given by \be{\mathcal
L}_{CS}=\f12\varepsilon^{mnl}(f^{ab\bar{c}\bar{d}}A_{m\bar{c}b}\partial_n
A_{l\bar{d}a}+\f23f^{ac\bar{d}}_{\ \ \ g}
f^{ge\bar{f}\bar{b}}A_{m\bar{b}a}A_{n\bar{d}c}A_{l\bar{f}e}).\ee

Note that the Lagrangian (\ref{Lagr6}) is automatically gauge
invariant since it is super\-sym\-met\-ric and super\-sym\-met\-ries
close into gauge transformations \be\label{closeZ}
[\delta_1,\delta_2]Z^A_d=v^m\cD_m
Z_d^A+\Lambda_{\bar{c}b}f^{ab\bar{c}}_{\ \ \ d}Z^A_b,\ee where \be
v^m=\f{i}{2}\epsilon_2^{CD}\gamma^m\epsilon_{1CD}, \quad
\Lambda_{\bar{c}b}=i(\bar\epsilon^{DE_2}\epsilon_{1CE}-\bar\epsilon_1^{DE}\epsilon_{2CE})\bar{Z}_{D\bar{c}}Z^A_b.\ee
The second term in (\ref{closeZ}) is a gauge transformation:
$\delta_\Lambda Z^A_d= \tilde{\Lambda}^a_dZ^A_a$. On the field
$\bar{Z}_{A\bar{d}}$ we find
$\delta_\Lambda\bar{Z}_{A\bar{d}}=\Lambda^\star_{c\bar{b}}f^{\star\bar{a}\bar{b}c}_{\
\ \ \bar{d}}\bar{Z}_{A\bar{a}}$. If we assume the existence of a
gauge invariant metric, namely
$\delta_\Lambda(h^{\bar{a}b}\bar{Z}_{A\bar{a}}Z^A_b)=0,$ we must
require \be f^{ab\bar{c}\bar{d}}=f^{ab\bar{c}}_{\ \ \
e}h^{\bar{d}e}=f^{\star\bar{c}\bar{d}ab}.\ee This implies that
$(\tilde{\Lambda}^{c\bar{d}})^\star=-\tilde{\Lambda}^{d\bar{c}},$
therefore the transformation parameters $\tilde{\Lambda}^a_b$ are
elements of $u(N)$. In this example we see that the general form of
three-dimensional Lagran\-gians with $\cN=6$ supersymmetry, $SU(4)$
R-symmetry and a $U(1)$ global symmetry with gauge group $U(N)\times
U(N)$ is entirely determined by specifying a triple product on a
3-algebra that satisfies the fundamental identity. It would
certainly be interes\-ting to see if there are other examples and
hence other models with different gauge groups. A matrix realization
of the Hermitian 3-algebra \cite{BL-N6} is proved by
$[X,Y;Z]=XZ^\dag Y-YZ^\dag X$, $<X,Y>=\mbox{tr}(XY^\dag).$ The
matrix-value fields $X,Y,Z$ are expanded as $X=X_aT^a$ etc., where
$T^a$ is a basis of $(M\times N)$ matrices and $T_a$ are their
Hermitian conjugates. The 3-bracket is then a map from $M\times N$
matrices to itself as the first requirement of an algebra. Moreover,
the bracket satisfies the fundamental identity (\ref{fi-N6}). Hence,
it is a realization of the Hermitian 3-algebra. An explicit solution
of the fundamental identity can also be realized in terms of the
generators $t^\alpha$ of the associated semi-simple Lie algebra as
\be f^{ab}_{\ \ \ cd}= (t^\alpha)^{a}_{\  d}(t_\alpha)^b_{\
c}~,\ee where $(t^\alpha)^{a}_{\  b}$ are the generators in the
bi-fundamental representation. The index $\alpha$ is lowered by the
inverse of Killing form $\kappa^{\alpha\beta}$ of the Lie algebra.
This realization does not in general satisfy antisymmetry with
respect to $a,b$ or $c,d$ indices. Imposing this property restricts
possible choices of the Lie algebras and hence the Lie group. With
the Lie group ${\mathcal G}=G_L\otimes G_R$, $a,b,c,d$ ranges over
$1,\ldots,\mbox{rank}(G_L)\mbox{rank}(G_R)$ and $\alpha$ ranges over
$1,\ldots,\mbox{rank}(G_L)+\mbox{rank}(G_R).$ As shown in works
\cite {schnabl} after analysis of all possible compact Lie groups
and their representations, only allowed gauge groups leading to the
manifest $\cN = 6$ supersymmetry are, up to discrete quotients,
$SU(N)\times U(1)$, $Sp(N)\times U(1)$, $SU(N) \times SU(N)$, and
$SU(N) \times SU(M) \times U(1)$ with possibly additional $U(1)$’s.
Matter representations are restricted to be the bi-fundamentals. But
we have to emphasize the role of triple products and 3-algebras even
though the resulting Lagrangians can be viewed as relatively
familiar Chern-Simons-matter gauge theories based on Lie algebras.
From the point of view declared here, the dynamical fields have
interactions that are most naturally defined in terms of a triple
product.

After the proposal of BLG, a lot of attempts have been done to
extract and understand various aspects of this theory. One of the
important articles in this direction is the paper of \cite{mukhi} in
which it was shown that if one of the scalars, for example $X^8$,
has a nonzero expectation value, one can reduce the membrane action
to D2 brane action which shows an important notion of reliability of
the BLG theory.

In a very interesting series of publications \cite{hohmN}, the
authors proposed to approach the construction of three-dimensional
superconformal gauge theories for all values of $\cN$ by making use
of a relation with gauged supergravity. Three-dimensional
supergravity theories differ from their higher-dimensional relatives
in that all bosonic degrees of freedom can be described by scalar
fields. These can be seen as coordinates of a manifold, on which
supersymmetry imposes a number of geometric conditions. For $\cN >
4$ these are strong enough to completely fix the (ungauged) theory:
the scalar manifolds are given by certain symmetric spaces. The
vector fields needed for the gauging only occur inside the covariant
derivatives and via a Chern-Simons term but do not have a kinetic
term. Their field equations lead to a duality relation between the
vectors and the scalars such that no new degrees of freedom are
introduced. This method was originally developed in the construction
of maximal $\cN = 16$ supergravities \cite{nicolai}, where the most
general $\cN = 16$ gaugings encoded in the 'embedding tensor' were
classified. The role of this tensor is to specify which subgroup of
the global symmetry group of isometries a manifold bosonic degrees
of freedom is gauged and which vectors are needed to perform this
gauging. In conformal limit upon sending Newton's constant to zero,
the supergravity and matter multiplets decouple. The resulting
theory for the matter multiplets has $\cN$ global supersymmetries.
In supergravity there is a number of restrictions on which
transformations can be gauged. These can be succinctly summarised in
terms of a linear and a quadratic constraint on the embedding
tensor. The quadratic constraint follows from the requirement that
the embedding tensor itself is invariant under the transformations
that are gauged. The linear constraint on the embedding tensor
follows from supersymmetry. In other words, it is perfectly
consistent to introduce gaugings that do not satisfy the linear
constraint, but these will not preserve supersymmetry. As it follows
from the requirement of supersymmetry, this condition takes a
different form for different values of $\cN$. In \cite{hohmN} the
authors present a systematic way to solve these constraints, which
reproduces the classification of superconformal theories for
different values of $\cN$ given in the recent literature \cite{BLG},
\cite{no-go} -\cite{mukhi}. They also find three new superconformal
theories with $\cN=4,5$ supersymmetry. One advantage of the
supergravity approach is that the same idea can be used to obtain
non-conformal theories as well by taking other limits.

In \cite{BILPSZ} it have been constructed the classical action of
the ABJM model in the $\cN=3$, $d3$ harmonic superspace
\footnote{For the $\cN=1,2$ superspace formulations of the BLG and
ABJM model see\cite{n12}}. Our motivation comes in part from
corresponding studies in $AdS_5/CFT_4$ where the $\cN=2$ formulation
of $\cN=4$ supersymmetric Yang-Mills theory (SYM) has been an
extremely efficient tool for studies of anomalous dimensions,
non-renormalization properties and integrability. In such a
formulation three out of six supersymmetries are realized off shell
while the other three mix the superfields and close on shell. The
superfield action involves two hypermultiplet superfields in the
bifundamental representation of the gauge group and two Chern-Simons
gauge superfields corres\-pon\-ding to the left and right gauge
groups. The $\cN=3$ superconformal invariance allows only  a minimal
gauge interaction of the hypermultiplets. One may wonder how the
sextic scalar potential of the ABJM model can appear in the absence
of an original superpotential. We show that, upon reducing the
superfield action to the component form, the scalar potential
naturally arises as a result of eliminating some auxiliary fields
from the gauge multiplet and from the harmonic expansion of the
off-shell $q^+$ hypermultiplets. This is a striking new feature of
the $\cN=3$ superfield formulation as compared to the $\cN=1$ and
$\cN=2$ ones. Besides the original $U(N)\times U(N)$ ABJM model, we
also constructed $\cN=3$ superfield formulations of some
generalizations. For the $SU(2) \times SU(2)$ case we give a simple
superfield proof of its enhanced $\cN=8$ supersymmetry and $SO(8)$
R-symmetry. To clarify the significance of the $\cN=3$ superfield
formulation presented in \cite{BILPSZ}, let us resort to the analogy
between the ABJM theory and the $\cN=4$, d4 super Yang-Mills
(SYM$^4_4$) theory, which describe the low-energy dynamics of
multiple M2 and D3 branes, respectively. It is well known, the
SYM$^4_4$ model is the maximally supersymmetric and superconformal
gauge theory in four dimensions, a fact is crucial for the string
theory / field theory correspondence (see e.g. \cite{maldac}). The
$\cN=2$, d4 harmonic superspace \cite{gios} provides the appropriate
off-shell $\cN=2$ superfield description of SYM$^4_4$ as SYM$^2_4$
plus an $\cN=2$ hypermultiplet in the adjoint representation
minimally coupled to the $\cN=2$ gauge superfield. Such a
formulation was successfully used to study the low-energy quantum
effective action and the correlation functions of composite
operators in $\cN=2$ superspace. Analogously to SYM$^4_4$, the ABJM
model is the maximally supersymmetric and superconformal
Chern-Simons-matter theory in three dimensions. The ABJM
construction opened up ways for studying the AdS$_4$/\-CFT$_3$
correspondence between three-dimensional field models and
four-dimensional super\-gra\-vity in AdS space \cite{ABJM}. We
believe that the $\cN=3$ superfield description of the ABJM model
and its generalizations developed in the paper \cite{BILPSZ} will be
useful for studying their algebraic and quantum structure as the
$\cN=2$ harmonic superspace approach has proved to be for SYM$^4_4$.
In particular, we expect that it will be very efficient for
investigating the low-energy quantum effective action in
three-dimensional $\cN=6$ supersymmetric field models superspace,
because the manifest off-shell $\cN=3$ supersymmetry is respected at
each step of the computation. One of the most interesting features
of the BLG model is the subtle interplay between the gauge algebra
and supersymmetry, and we hope that our manifestly supersymmetric
formulations will shed more light on this issue.

\section{M-brane bound states and the supersymmetry of BPS
solutions in the BLG theory}

If the BLG theory is  provided an authentic description of multiple
M2-branes, it must be able to incorporate the various M-branes which
are known to exist. They are supersymmetric objects of the 11
dimensional quantum supergravity, and will appear as classical BPS
solutions in the dual field theory, the BLG Lagrangian.  They imply
the existence of M-branes in addition to the 'background M2-branes'
whose dynamics is describe by the BLG theory in question. Since in
$\cM$-theory for an M2 brane there is an M5 brane which is
electric-magnetic dual of each other, one natural task is to find
the relation between the M2 and M5 brane dynamics in the context of
the BLG theory. In the case of string theory, such solutions have
been written down explicitly for the case where D1-branes expand
into a single D3-brane  and into multiple intersecting D3-branes.
M2-branes can blow up into BPS funnels that end on calibrated
intersections of M5-branes. In \cite{krish} the authors make the
observation that the constraints required for the consistency of
these BPS solutions are automatic in BLG theory, thanks to the
fundamental identity and the supersymmetry of the calibration.

The original motivation of Bagger and Lambert was to write down a
theory capable of reproducing the Basu--Harvey equation \cite{basu}
that describes an M2-brane ending on an M5-brane as a BPS equation.
This generalized Nahm's equation \cite{nahm}, usually called ADHM
construction, for the moduli space of monopoles in the gauge theory
which describes a D1-brane ending on a D3-brane. Nontrivial BPS
solutions would have less supersymmetries and they disintegrate into
three categories: the vortices, the domain walls and the
spacetime-filling configurations. Simple 1/2-BPS equations can be
readily written and also the solutions have been studied, see
\cite{BLG}. The energy bound corresponding to this particular BPS
configuration should appear in the superalgebra of the theory as a
central charge term.

When we compute the anticommutator of two supercharges, we obtain
the follo\-wing result: \be \{Q^\alpha,
Q^\beta\}=-2P_m(\Gamma^m\Gamma^0)^{\alpha\beta}+Z_{IJ}(\Gamma^{IJ}\Gamma^0)^{\alpha\beta}+Z_{iIJKL}(\Gamma^{IJKL}\Gamma^i\Gamma^0)^{\alpha\beta}\ee
$$+Z_{IJKL}(\Gamma^{IJKL})^{\alpha\beta},$$
where $\alpha, \beta$ are the 11 dimensional spinor index, and
$i=x,y.$ In the above we have, in addition to the usual energy
momentum vector $P_m$ defined as $P^m=\int d^2x T^{0m}$, three types
of central charges: \be Z_{IJ}=-\int d^2x \mbox{tr}(\cD_i X^I\cD_j
X^J \varepsilon^{ij}-\cD_0X^K F^{KIJ}),\ee
$$Z_{iIJKL}=\f13\int d^2x \mbox{tr}(\cD_j X^{[I}F^{JKL]} \varepsilon^{ij} ),$$
$$Z_{IJKL}=\f14\int d^2x \mbox{tr}(F^{M[IJ}F_M^{KL]}  ),$$
where we  also introduced a short-hand notation for 3-products:
$F^{IJK}=[X^I,X^J,X^K].$ The stress-energy tensor $T_{mn}$ can be
computed in the usual way. In case where the fermions are set to
zero, it results in \be
T_{mn}=\cD_mX^I_a\cD_nX^{aI}-\eta_{mn}(\f12\cD_\rho X^{aI}\cD^\rho
X^I_a +V)~.\ee We note that the Chern--Simons like term does not
contribute to the stress-energy tensor because this term is
topological one and does not depend on the worldvolume metric. The
first two classes are actually topological terms, since they can be
expressed as surface integrals. They are boundary terms and they are
equal to zero for field configurations that are non-singular and
topologically trivial. The last one, $Z_{IJKL}=Z_{[IJKL]}$ can be
actually shown to vanish as well, but for a different reason: one
should make use of the invariance, the fundamental identity and
skew-symmetry.

Now we are almost ready to consider simple BPS equations and
identify the central charge terms as different combinations of
M-branes. We consider vortex configurations that describe two stacks
of membranes intersecting along the time direction where only the
scalars $X^3=\Phi+\bar\Phi, X^4=i(\Phi-\bar\Phi)$ and the gauge
vector $\tilde{A}_{na}^b$ are excited. Thus, considering a
configuration such that $\cD_0 \Phi=\cD_0\bar\Phi$, the BPS
conditions that follow from supersymmetry variations
(\ref{susytrans}) are reduced to
\footnote{$\Gamma^z=\Gamma^1+i\Gamma^2, \quad
\Gamma^{\bar{z}}=\Gamma^1-i\Gamma^2, \quad
\Gamma^\Phi=\Gamma^3+i\Gamma^4, \quad
\Gamma^{\bar\Phi}=\Gamma^3-i\Gamma^4$}
\be\label{vortBPS}\cD_z\Phi\Gamma^z\Gamma^\Phi\epsilon+\cD_{\bar{z}}\Phi\Gamma^{\bar{z}}\Gamma^\Phi\epsilon+\cD_z\bar\Phi\Gamma^z\Gamma^{\bar\Phi}\epsilon+
\cD_{\bar{z}}\bar\Phi\Gamma^{\bar{z}}\Gamma^{\bar\Phi}\epsilon=0.
\ee For this configuration, the energy density is given by
$${\mathcal H}=4\mbox{tr}(\cD_z\Phi, \cD_{\bar{z}}\bar\Phi)+4\mbox{tr}(\cD_{\bar{z}}\Phi, \cD_{z}\bar\Phi)=\f12 {\mathcal Z}^0+
8\mbox{tr}(\cD_{\bar{z}}\Phi, \cD_{z}\bar\Phi),$$ where ${\mathcal
Z}^0$ is the density of the 0-form central charge $Z_{IJ}$ evaluated
for this field configuration. Thus ${\mathcal H}\geq \f12{\mathcal
Z}^0$ and the bound is saturated when BPS configuration is given by
(anti)holomorphic curves
$$\cD_{\bar{z}}\Phi=\cD_z\bar\Phi=0.$$
If this last condition is satisfied, it follows from the BPS
equation (\ref{vortBPS}) that the solution preserve half of the
supersymmetries (\ref{susytrans}) satisfying
$\Gamma^z\Gamma^\Phi\epsilon=0$. Thus, for the case when the gauge
field is equal to zero, the vortex configuration is given by
$$\Phi=c_aT^a\f{1}{z},$$
where $c_a$ are arbitrary constants. For the case of the
configuration when also the gauge vector $A_m^{\tilde{a}}
T^{\tilde{a}}$  exists, we find that half-BPS exist if
$$[\Phi, A_{\bar{z}}]=[\bar\Phi, A_z]=0,$$ where $[\cdot,\cdot]$ is the usual Lie commutator. In this model, the 3-algebra indices $a$ are split into $a=(+,-,\tilde{a})$ and the structure
constants are given by
$$f^{+\tilde{a}\tilde{b}\tilde{c}}=f^{-\tilde{a}\tilde{b}\tilde{c}}=C^{\tilde{a}\tilde{b}\tilde{c}},
\quad
f^{+-\tilde{a}\tilde{b}}=f^{\tilde{a}\tilde{b}\tilde{c}\tilde{d}}=0,
$$
where $C^{\tilde{a}\tilde{b}\tilde{c}}$ are the structure constants
of a compact semi-simple Lie algebra satis\-fying the usual Jacobi
identity. This implies that with respect to the single M2-brane
theory, the vortex solutions of the BLG theory include extra degrees
of freedom, given by the components of the gauge vector that commute
with the scalar fields.

To describe a stack of M2-branes ending on an M5-brane it is
necessary to switch on the $X^3,X^4,X^5,X^6$ scalar fields
\cite{basu}. Given that these fields depend only on the worldvolume
coordinate $\sigma^2$, the BPS condition is \cite{BLG} \be\label{bps
BasH} \f{d
X^A}{d\sigma^2}\Gamma^A\Gamma^2\epsilon-\f16\varepsilon^{BCDA}\Gamma^A[X^B,X^C,X^D]\Gamma^{3456}\epsilon=0,\ee
where $A,...=3,4,5,6$. For this field configuration the energy
density is given by
$${\mathcal H}=\f12\mbox{tr}(\partial_2 X^A, \partial_2 X^A)+\f{1}{12}\mbox{tr}([X^A,X^B,X^C],[X^A,X^B,X^C]).$$
As usual, we can write the potential as
$V(X)=\f12\mbox{tr}(\f{\partial W}{\partial X^A}, \f{\partial
W}{\partial X^A})$, where \be W = \f12 m
\mbox{tr}(X^A,X^A)+\f{1}{24} \varepsilon^{ABCD} \mbox{tr}
(X^A,[X^B,X^C,X^D]).\ee Here we add an $SO(4)$ symmetric mass
deformation term. Thus
$${\mathcal H}=\f12\mbox{tr}(\partial_2 X^A+\f{\partial W}{\partial X^A},\partial_2 X^A+\f{\partial W}{\partial X^A})-
\mbox{tr}(\partial_2 X^A,\f{\partial W}{\partial X^A}),$$ where
${\mathcal Z}_1=-2\mbox{tr}(\partial_2 X^A,\f{\partial W}{\partial
X^A})$ is the density of $Z_i^{\alpha\beta}$ the 1-form central
charge. For this field configuration ${\mathcal H}\geq\f12 {\mathcal
Z}_1$ and the bound is saturated when \be\label{bas-harv} \f{d
X^A}{d\sigma^2}-\f16\varepsilon^{BCDA}[X^B,X^C,X^D]=m X^A. \ee When
the (\ref{bas-harv}) with $m=0$ are satisfied, then it follows from
(\ref{bps BasH}) that the field configuration proposed by Basu and
Harvey as the M2-brane worldvolume solution describing the M2-M5
system is half-BPS and the preserved supersymmetries satisfy
$\Gamma^2\epsilon=\Gamma^{3456}\epsilon.$

Vacuum solutions require $\partial_A W=0,$ or $m
X^A=-\f16\varepsilon^{BCDA}[X^B,X^C,X^D],$ where the $T^A$ satisfy
$[T^A,T^B,T^C]=\varepsilon^{ABCD}T^D.$ In addition to the trivial
solution $X^A=0,$ this Eq. has a fuzzy $S^3$ solution in which the
M2' puff up into a fuzzy three-sphere with radius proportional to
$\sqrt{m}$. This implies that $\mbox{tr}(T^A, T^A) \sim n^2$ at
large  integers $n$. Unfortunately, we do not know enough about the
representations of three-algebras to confirm this prediction. The
two solutions describe two zero-energy vacuum states of the M2-brane
in the four-flux background.

The fuzzy funnel solution of Eq. (\ref{bas-harv}) can be found by
taking \be X^A=f(\sigma^2)T^A .\ee The equation for $f$ is $f'=m
f-f^3;$ the solution is
$$f=\f{\sqrt{m}}{\sqrt{1-ce^{-2m\sigma^2}}}~.$$
If $c=+1$ and $m>0$ the solution behaves as
$f=\f{1}{\sqrt{\sigma^2}}$ for small but positive $\sigma^2.$ These
solutions describe fuzzy funnels in which an infinite radius fuzzy
three-sphere at $\sigma^2 = 0$ relaxes into the fuzzy sphere or the
trivial vacuum, respectively, as $\sigma^2 \rightarrow \infty.$ The
spacetime interpretation of these solutions is that they correspond
to M2-branes that end on a single M5-brane, located at $\sigma^2 =
0$ and infinitely extended along the $(x^0,\ldots, x^5)$ directions.
On the other hand, if $c =-1$ and $m > 0$, the function $f$ is
bounded. It vanishes exponentially as $\sigma^2\rightarrow -\infty $
and approaches $f \rightarrow \sqrt{m}$ as $\sigma^2 \rightarrow
\infty $. Here there is no divergent fuzzy funnel, i.e. no M5-brane.
This solution smoothly interpolates between the trivial and fuzzy
sphere vacua. In other words, it is a traditional domain wall that
interpolates between two degenerate vacuum solutions of the
worldvolume effective action.

Here we have only considered the simplest solutions and it would be
interesting to systematically work out more solutions and study
their properties. In the papers \cite{Hai} was studied two problems
of M5-branes in the $\cN = 6$ theory. The authors analyzed the
Basu-Harvey type equations and found evidence that the equations
describe multiple M2-branes ending on a M5-brane, which wraps on a
fuzzy 3-sphere.  They also derived the Nahm equation describing
D2-branes ending on a D4-brane wrapping an $S^2$ starting from the
Basu-Harvey type equations and taking a large k limit, providing
further evidence for consistency.  Then they turned to another
situation where M5-branes wrapping on fuzzy 3-sphere emerge as the
vacua of the mass-deformed $\cN = 6$ theory.

\section{Dynamical Symmetry and Nambu Mechanics}
While very little is known about explicit nontrivial examples of the
$n$-algebra, its correspondence with Nambu brackets is very helpful.

In this section we review some standard facts about the Nambu
brackets and $n$-algebras (Filippov--Nambu algebras). More than
three decades ago, Nambu \cite{Nambu} proposed a generalization the
classical Hamiltonian mechanics. Dynamics according to Nambu
consists in replacing Poisson bracket by a ternary (n-ary) operation
on algebra of observables $A$ and requires two ($n-1$)
'Hamiltonians' $H_1,H_2$ ($H_1,\ldots,H_{n-1}$) to describe the
evolution. This dynamical picture is consistent if and only if the
evolution operator is an isomorphism of algebra of observables.

This mechanics is remarkable in several respect. First, it treats
all conserved quantities of a mechanical system on the same footing.
It is clearly a most attractive feature from a quantum perspective.
In his formalism, Nambu replaced the usual pair of canonical
variables found in Hamiltonian mechanics with a triplet of
coordi\-na\-tes in an odd dimensional 'phase space' possessing as
fundamental symmetries the volume preserving diffeomorphisms group
in the place of symplectic diffeomorphisms. It has been shown that
several Hamiltonian systems possessing dynamical or hidden
symmetries can be realized within the framework of Nambu's
generalized mechanics. Among such systems are the Euler equations
for the angular momentum of a rigid body in three dimensions, the
$SU(n)$-isotropic harmonic oscillator, the $SO(4)$-Kepler problem
and others somewhat exotic examples. As required by the formulation
of the Nambu dynamics, the integrals of motion needed for complete
integrability of these systems necessarily become the so-called
generalized 'Hamiltonians'. Corres\-pon\-ding phase flow preserves
the phase volume so that the analog of the Liouville theorem is
still valid, which is fundamental for the formulation of statistical
mechanics with two temperature-like intensive parameters. Yet, the
fundamental principles of a canonical formalism of Nambu's
generalized mechanics and on the principle of least action, similar
to the invariant geometrical form of the Hamiltotian mechanics
\cite{arnold}, has only recently been given an elegant geometric
formulation by Takhtajan \cite{takh}. For further applications of
the classical and quantum Nambu brackets the reader may consult
\cite{zachos}.

The basic properties of the associative algebra
$\cA=C^\infty(\mathbb{R}^n)$, what equipped with a Nambu bracket of
order $n$ $\{\cdot, \ldots,\cdot\}$ the same as in $n$-Lie algebra
Filippov is: Linearity, Antisymmetry, Leibnitz rule and the
Fundamental Identity \be\label{FINambu}
\{\{f_1,\ldots,f_n\},f_{n+1},\ldots,
f_{2n-1}\}=\{\{f_1,f_{n+1}\ldots,f_{2n-1}\},f_{2},\ldots, f_{n}\}+
\ee
$$+\{f_1,\{f_2,f_{n+1}\ldots,f_{2n-1}\},f_{3},\ldots, f_{n}\}+\ldots+\{f_1,,\ldots,f_{n-1},\{f_{n},f_{n+1},\ldots, f_{2n-1}\}\}~.$$
On the vector space $V$ the linear Nambu brackets are related to the
notion of Filippov--Nambu $n$-algebra. If we introduce the basis
$T^a$ of $V$ then $n$-bracket can be defined through the structure
constants (\ref{struct_const}) and fundamental identity (\ref{FI}).

The identity (\ref{FINambu}) guarantees the fact that if each of
$(f_i)_{ i=1, \ldots, n}$ is the  conserved quantity, then the
observable $\{f_1, \ldots, f_n\}$ is also conserved. The new
equations of motion in the phase space $M\equiv \mathbb{R}^n$ are
analogous to the Hamilton--Poisson equations: \be \f{d
x^i}{dt}=\{x^i, H_1,\ldots,H_{n-1}\},\ee where the $n$-bracket on an
oriented $n$-dimensional manifold is defined as Jacobian (volume
form): \be\label{nambr} \{f_1,\ldots,f_n\}=\varepsilon^{i_1\ldots
i_n}\partial_{i_1}f_1\partial_{i_2}f_2\ldots\partial_{i_n}f_n~,\ee
for any functions $f_1,\ldots,f_n \in C^{\infty}(\mathbb{R}^n)$ and
$i=1,\ldots,n$. The $n-1$ Hamiltonians $H_1, \ldots, H_{n-1}$
determine the phase-space trajectory in a geometrical way. These is
also a corresponding Liouville equation for any observable $f\in
C^{\infty}(\mathbb{R}^n)$ \be\f{df}{dt}=\partial_if\cdot
\dot{x}^i=\{f,H_1, \ldots,H_{n-1}\}.\ee The $n-1$ Hamiltonians are
conserved in time. Given the initial position in the phase-space
$x_0^i=x^i(t=0)$ they take the values $h_i=H_i(x_0)$. The
intersection of hypersurfaces $h_i; i=1,\ldots, n-1$ gives the
geometrical shape of the trajectory passing through the point $x_0
\in \mathbb{R}^n$. This is the reason why the Nambu 3-d dynamical
system is regarded as a toy model for completely integrable systems.
To make a case for the physical relevance of this new formalism,
Nambu pointed out a specific realization for the $n=3$; namely the
asymmetric Euler top. Here the triplet $\vec{x}$ is naturally
identified with the angular momentum $\vec{l}$ in the body fixed
frame. There are two guaranteed conserved quantities: the total
kinetic energy
$H_1=\f12(\f{l_1^2}{I_1}+\f{l_2^2}{I_2}+\f{l_3^2}{I_3})$ and
$H_2=\f12(\vec{l})^2$ the Casimir invariant. The corresponding phase
space is $S^2$ which provides a spherical foliation of
$\mathbb{R}^3$ with varying radius $\sqrt{2H_2}$ for various
conditions $l_0^i$ with Poisson algebra $SO(3)$:
$\{l^i,l^j\}=\varepsilon^{ijk}l^k$. The classical Nambu Eqs. of
motion are $\dot{l}^i=\varepsilon^{ijk}\partial_jH_1\partial_kH_2$
or \be\label{eq top}\dot{l}_1=(\f{1}{I_2}-\f{1}{I_3})l_2l_3, \quad
\dot{l}_2=(\f{1}{I_3}-\f{1}{I_1})l_3l_1,\quad
\dot{l}_3=(\f{1}{I_1}-\f{1}{I_2})l_1l_2~,\ee which are just the
Euler force-free rigid body equations. In this example the time
evolution of the Euler top in Nambu mechanics is described by two
Hamiltonian functions. These two Hamiltonians lie in the same
$SO(3)$ Lie algebra \footnote{This situation contrasts with Dirac's
mechanics where the constraints appear as subsidiary conditions} and
are interpreted, the first one as the one who defines the 2-d phase
space geometry, embedded in the 3-d phase space, while the second
one gives the dynamics of the trajectories on the 2-d phase space.
Specifically the Euler equations for the asymmetric top naturally
describe geodesic flows on a triaxial ellipsoid and can be solved in
terms of Jacobi elliptic function \cite{landau}. It should be noted
that these equations have reincarnated during recent decades in the
celebrated Nahm equations \cite{nahm} for the SU(2) self-dual
Yang--Mills field relevant to theories of extended objects such as
monopoles and membranes.

In addition it is important  that we can lay the basis for a
generalized quantum mechanics based on the classical Nambu dynamics
\cite{minic} of Euler’s asymmetric top. This Nambu quantum mechanics
naturally possesses, besides Planck constant, new deformation
parameters. One of its defining experimental signatures is a
nonlinear time evolution generated by Jacobi
 elliptic functions, as
compared to the standard exponential time evolution of standard
quantum mechanics. The new deformation parameters are given by the
moduli of the elliptic functions. In the limit when these are set to
zero, the usual geometric formulation of quantum mechanics, based on
the K\"ahler structure of the space of rays in a complex Hilbert
space, is recovered. This motivates the general expression for what
we call the Nambu wave function $\Psi^a=\sum_i l^a_i(t) e_i$ ($a=1,
\ldots, N$) where $e_i$ are the usual quaternion imaginary units
such that $e_ie_j = -\delta_{ij} + \varepsilon_{ijk}e_k$. The
quaternion conjugate Nambu wave function is
$\bar\Psi^a=-\sum_il^a_i\bar{e}_i$. The inner product reads
$\bar\Psi\Phi=\delta_{ij}\Psi_i\Phi_j-\varepsilon_{ijk}e_k(\vec{\Psi}\times\vec{\Phi}).$
The second term in the above equation is the quaternionic
counterpart of the symplectic 2-form. It is at the basis of the
3-form, characteristic of Nambu's original mechanics \cite{Nambu}.
Due to the non-linear nature of the Nambu--Schr\"odinger equation
that describes a collection of $N$ (which could be infinite) free
abstract Euler tops the superposition principle apparently no longer
holds for $\Psi$.

Next the fundamental analog of the symplectic 2-form of usual
Hamiltonian dynamics is a closed non-degenerate 3-form \cite{takh}:
$\omega^{(3)}=d l_1\wedge dl_2 \wedge dl_3$, and the action given as
an integral of the corresponding Poincare--Cartan 2-form $S=\int l_1
dl_2 \wedge dl_3-H_1 dH_2\wedge dt.$ This form of the action shows
that initial and final states in this type of the Nambu dynamics are
described by loops rather than points, because the integrand of the
action is a two form, rather than a one form, as in the usual
Hamiltonian dynamics. Now, we can say that, just as the simple
harmonic oscillator is the prototype classical and quantum system of
the standard Hamiltonian mechanics, the Euler asymmetric top is the
prototypical representative of Nambu's ternary mechanics.

 It is straightforward to generalize presented results for the case of the Nambu bracket of order $n$.
 The analog of the Poincare-Cartan integral invariant is defined as the following $n-1$
 form
\be \omega^{(n-1)}=x_1 dx_2\wedge\ldots\wedge dx_n-H_1 dH_2\wedge
\ldots \wedge dH_{n-1}\wedge dt .\ee
 The action functional is given by
 \be S(C_{n-1})=\int_{C_{n-1}}\omega^{(n-1)} \ee
 and is defined on the $n-1$-chains in the extended phase space. In its formulation admissible variations are those which do not change projections of the boundary
 $\partial C_{n-1}$
 on the $x_2 x_3 \ldots x_n$-hyperplanes; in this case the "share" of "configuration space" in a phase space is $1-\f{1}{n}$.

 Next we consider the all-familiar
 classical Coulomb problem $\f{dz^i}{dt}=\{ z^i, H \},$ with $z^i$ standing for the phase-space 6-vector $(\vec{r},\vec{p}),$ and $H=\f{p^2}{2}-\f{1}{r}.$
 In discussing of this example as an illustration of the
 general method we follow the original publications \cite{zachos04}
 and the references therein.
 Because $H$
 possesses rotational symmetry, the orbital angular momentum $\vec{L}=\vec{r}\times \vec{p}$ is an integral of
 motion.
 This rotational symmetry implies that the orbit
 lies in some
 two dimensional plane, though it is not enough to ensure that the orbit is closed. An extra dynamical symmetry must exist for closed orbit. Such an integral was
 first discovered by Laplace (but is called the Runge--Lenz vector in classical mechanics or the Lenz--Pauli vector in quantum mechanics) and is given by
 $\vec{A}=\vec{p}\times\vec{L}
 -\f{\vec{r}}{r}.$ Multiplying it by $\vec{n}=\f{\vec{r}}{r}$ instantly yields Kepler's elliptical orbits, $\vec{n}\cdot \vec{A}+1=\vec{L}^2/r$.
 Since $\vec{A}\cdot\vec{L}=0,$ it
 follows that $H=\f{A^2-1}{2L^2}$. One can easily check that $\{A_i,L_j\}=\varepsilon_{ijk}A_k$ and $\{A_i,A_j\}=-2HL_k.$ For bound state problems $(E<0)$, one can define a new conserved
 vector $\vec{D}=\f{\vec{A}}{\sqrt{-2E}}$ and further ${\mathcal R}\equiv \vec{L}+\vec{D}$, ${\mathcal L}\equiv \vec{L}-\vec{D}$. These six simplified invariants obey the standard
 $SU(2)\times SU(2)\sim SO(4)$ symmetry algebra (Note that for scattering problems where $E>0$, one instead find the Lorentzian Lie algebra $SO(3,1)$),
 $$\{{\mathcal R}_i,{\mathcal R}_j\}=\varepsilon_{ijk}{\mathcal R}_k,\quad
 \{{\mathcal R}_i,{\mathcal L}_j\}=0, \quad \{{\mathcal L}_i,{\mathcal L}_j\}=\varepsilon_{ijk}{\mathcal L}_k$$
 and depend on each other and the Hamiltonian through $H=\f{-1}{2{\mathcal R}^2}=\f{-1}{2{\mathcal L}^2}$, so
 only five of the invariants are algebraically independent. Equivalently
 to the Hamiltonian law of motion $\f{dz^i}{dt}=\{z^i, H\}$ the same classical evolution may also be
 specified by Nambu's equation of motion that is the case for all superintegrable
 systems with five of the above six $L_i, A_i$ (or products thereof) as the generalized Hamiltonians
 $$\f{dz^i}{dt}=\f{1}{L_1 \cdot (L_1^2+L_2^2+L_3^2)}\f{\partial(z^i, A_2,A_3,L_1,L_2,L_3)}{\partial(p_1,p_2,p_3,r_1,r_2,r_3)},$$
 or
 $$\f{dz^i}{dt}=H^2\{z^i, \ln({\mathcal R}_3+{\mathcal L}_3),{\mathcal R}_1, {\mathcal R}_2,{\mathcal L}_1,{\mathcal L}_2\},$$
 etc.

So we feel that these new examples may help in further understanding
of the elements of the Nambu's theory such as its algebraic
structure and its possible quantiza\-tion \cite{zachos04}. Then we
need to understand in every specifically case which properties are
essential from the physical viewpoint. As noted by Pauli, extension
to operators requires a hermitian version of $\vec{A}$:
$\vec{A}'=\f12(\vec{p}\times\vec{L}-\vec{L}\times\vec{p})-\f{\vec{r}}{r}$
so that $(\vec{A}')^2= 2H(\vec{L}^2+\hbar^2)+1$, leading to ${\bf
D}'=\f{{\bf A}'}{\sqrt{-2H}}$ and further to the respective chiral
reduction $\mathcal R'$ and $\mathcal L'$, which obey
$$[{\mathcal R}'_i,{\mathcal R}'_j]=2i\hbar\varepsilon_{ijk}{\mathcal R}'_k,
\quad [{\mathcal R}'_i,{\mathcal L}'_j]=0, \quad [{\mathcal
L}'_i,{\mathcal L}'_j]= 2i\hbar\varepsilon_{ijk}{\mathcal L}'_k$$
and hence \be H=\f{-1}{2({\mathcal
R'}^2+\hbar^2)}=\f{-1}{2({\mathcal L'}^2+\hbar^2)}~.\ee We can now
recall the eigenvalues of the quadratic Casimir invariants of
$SU(2)$ for $s=0,\f12,1, \ldots$ leading to the Balmer spectrum for
the Hamiltonian $E=\f{-1}{2\hbar^2(2s+1)^2}=\f{-1}{2\hbar^2n^2}$.
The size of these $SU(2)\times SU(2)$ multiplets $(2s+1)^2=n^2$ is
the corresponding degeneracy. In this example, the Nambu brackets
quantization coincides with the standard Hamiltonian quantization.
Besides, the authors \cite{zach02} also note that the quantum Nambu
brackets are the good guide for investigation of more general
systems.

Now we will present the Lie algebra of volume preserving
diffeomorphisms $SDiff$ $(R^3)$ in the Clebsch--Monge gauge, their
relation with the Filippov--Nambu 3-algebras on $\mathbb{R}^3$ as
well as on $\mathbb{T}^3$ and the Nambu mechanics, which can be
represented as flow equations of incompressible fluids \cite{axe}.
Since the famous paper by V.Arnold \cite{arn66} where he proved that
the solution of the Euler Eqs. for perfect (incompressible and
inviscid ) fluids  are the geodesics of the infinite dimensional
volume preserving diffeomorphisms  group, there have been many
developments. Here we will focus in the description of $SDiff(R^3)$,
in a particular gauge, the Clebsch--Monge gauge, thus establishing
the connection with the Nambu dynamics (flows) in $\mathbb{R}^3$.
This discussion easily extends to three dimensional manifolds with a
metric and a smooth Nambu tensor field. Let ${\cA} = C^\infty(R^3)$
be the space of smooth functions on $\mathbb{R}^3$ and ${\mathcal G}
= SDiff(R^3)$ be the set of smooth maps of $\mathbb{R}^3 \rightarrow
\mathbb{R}^3$ with the determinant of the Jacobian at each point of
$\mathbb{R}^3$ equals to one, i.e. $J(f)(x) = \mbox{det}[\partial_i
f^i(x)] = 1.$ This set forms a group under composition of functions.
The elements $X(v)=-v^i\partial_i$ with the Lie algebra
$[X(u),X(v)]=X(w)$ for $f^i(x)=x^i+v^i(x)$ with $\partial_i v^i=0$
have composition law:
$w=(u\cdot\partial)v-(v\cdot\partial)u=\partial \times(u\times v).$
We will impose conditions at infinity for $v^i(x)$:
$v^i(x)\stackrel{|x|\rightarrow \infty}{\rightarrow} 0$ such that
the total kinetic energy is finite $E=\f12\int d^3x v^2(x)< \infty$.
For any infinitesimal element we define the flow
$\f{dx^i}{dt}=v^i(x)$ with initial conditions $x^i_0=x^i(t=0).$ This
Eq. describes the motion of a particle which is immersed in a fluid
of given stationary velocity field at the point $x_0^i$, at $t = 0$.
For every divergenceless vector field $v^i(x)\in \mathbb{R}^3$, with
above boundary conditions  we can find a vector potential $A^i(x)$
such that $v^i =\varepsilon^{ijk}\partial_jA_k.$ For such $A^i(x)$
Clebsch and Monge introduced three scalar potentials
$\alpha,\beta,\gamma \in C^\infty(R^3)$ such that:
$A_i=\partial_i\alpha+\beta\partial_i\gamma.$ So finally we get
$v^i(x) =\varepsilon^{ijk}\partial_j\beta\partial_k\gamma.$ The
scalar function $\alpha(x)$ becomes the gauge degree of freedom of
$A^i(x).$ From the last relation we see that the intersection of the
surfaces $\beta = \mbox{const}$,$\gamma = \mbox{const}$ define
locally the flow lines. The existence of the scalar potentials
$\beta, \gamma$ (Clebsch--Monge potentials) is guaranteed locally if
$v^i(x)$ is an analytic function in the region of a point say $x^i =
0$. Then there exists two integrals of motion of the flow equation:
$\f{dx^i}{v^i(x)}=dt$ through which we can determine $\beta$ and
$\gamma.$ The generators of the flow, in terms of the Clebsch--Monge
potentials, become $X(\beta,\gamma)  =
-\varepsilon^{ijk}\partial_j\beta\partial_k\gamma\partial_i$ and the
action of $X(\beta,\gamma)$ on a smooth function $\alpha \in
C^\infty(R^3)$ is: $X(\beta,\gamma)\alpha=-\{\alpha,\beta,\gamma\},$
the Nambu bracket of $\alpha,\beta,\gamma.$ The volume-preserving
property is nothing but the identity $\partial_i
X^i(\beta,\gamma)=\partial_k(\varepsilon^{ijk}\partial_i\beta\partial_j\gamma)=0.$
The flow  becomes $\dot{x}^i = \{x^i, \beta,\gamma\};$ and so the
Clebsch--Monge potentials of the flow are just the two Hamiltonians
$H_1=\beta$, $H_2=\gamma$ of the Nambu dynamics. We conclude that
the flow equations of incompressible fluids can be described by the
Nambu dynamics and vice versa. By considering now the commutation
relations of the algebra in the Clebsch--Monge gauge we obtain:
$[X(\beta_1,\gamma_1),
X(\beta_2\gamma_2)]=X(\{\beta_1,\gamma_1,\beta_2\},\gamma_2)+
X(\beta_2,\{\beta_1,\gamma_1,\gamma_2\}).$ Acting both sides of this
relations on functions $\alpha$ we get the fundamental identity:
$$\{\beta_1,\gamma_1\{\beta_2,\gamma_2,\alpha\}\}-
\{\beta_2,\gamma_2\{\beta_1,\gamma_1,\alpha\}\}=\{\{\beta_1,\gamma_1,\beta_2\},\gamma_2,\alpha\}\}
+\{\beta_2,\{\beta_1,\gamma_1,\gamma_2\},\alpha\}.$$ Thus we observe
that all the information of the commutation relations of
$SDiff(R^3)$ is contained in the Filippov-Nambu 3-algebra for a
basis of functions in $\mathbb{R}^3.$ So, if both Hamiltonians are
linear,  $H_1 = a\cdot x,$ $H_2 = b\cdot x$, then the flows $X(a, b)
= (a \times b)^i\partial_i$ represent translations along the
direction $a\times b$ (constant laminar flow). The next interesting
case is of the linear Nambu flow with an axis of symmetry $\vec{a}$,
which can be derived from a pair of Hamiltonians, $H_2 = \vec{a}
\cdot \vec{x} $ and $H_1 = 1/2(\vec{x}, B\vec{x})$, where $a, x \in
\mathbb{R}^3$ and $B$ is a real, symmetric, $3 \times 3$ matrix. The
corresponding trajectory of the linear Nambu flow  is given by
$\f{dx^i}{dt}= \varepsilon^{ijk}a^jB^{kl}x^l = x^lM^{li} .$ The
solutions, given an initial condition $x^i(0)$, lie on the
intersection of the plane with the normal vector $\vec{a}$ and the
quadratic surface given by $H_1 = 1/2(\vec{x}(0), B\vec{x}(0)).$ We
can integrate the equation of motion explicitly and find $\vec{x}(t)
= \vec{x}(0)e^{tM} ~.$ Since the matrix $M$ is traceless, $A = e^M$
is an element of the group $SL(3,R).$ It is possible to compactify
the linear Nambu flow on $\mathbb{T}^3$, if we consider the linear
evolution equation, modulo the size of the torus, i.e. we take $x^i$
to belong to the elementary cell, $x^i = x^i +L^i$, where $L^i$ is
the length of the torus along direction $x^i$. If we choose these
units so that $L^i = 2 \pi$ then the action of the matrix $A$ on
every point of $\mathbb{T}^3$ is then taken modulo $2 \pi$. These
flows are known \cite{arn66} as toral automorphisms. The motion in
this case, even though the equation is linear, can be chaotic,
depending on the matrix elements of $A.$ We can check that, for
linear Nambu flows in $\mathbb{R}^3$, we have, essentially, a
reduction to a two-dimensional phase space problem on the plane
orthogonal to the vector $\vec{a}$. In the case of $\mathbb{T}^3$,
if the vector has rational components, then we have a finite number
of different images of the plane; if, however, the components are
irrationals, then we have a truly three-dimensional evolution for
the system.

The other illustrative example for this construction is an electric
charge in a homogeneous magnetic field. At this point we review the
relationship between the noncommutativity and incompressibility of
the quantum Hall state and the effective theory of the
incompressible fluid for the quantum Hall effect that is one of the
most remarkable phenomena in condensed matter physics.

The classical phase space is defined by the $H_2$ function:
$H_2=\f{e}{m^2c}\vec{v}\cdot\vec{B}$ and so the Nambu algebra of the
phase-space coordinates $v^i$ is according to Eq.
$\{X^i,X^j\}_{H_2}=\epsilon^{ijk}\partial_kH_2$,
\be\{v^i,v^j\}=\f{e}{m^2c}\epsilon^{ijk}B_k.\ee The phase space is a
plane transverse to $B$ embedded in $R^3$. The dynamics is defined
through $H_1=\f12 mv^2$ and the Hambu Eqs
$$\dot{v}^i=\f{e}{mc}\epsilon^{ijk}v^jB^k$$ produce the correct physical
Eqs. of motion for the Landau problem. The density of states in the
lowest Landau level (LLL) is uniform and in proportion to the
strength of the magnetic field, $\rho_0=\f{1}{2\pi l_0^2},$ where
$l_0 = 1/\sqrt{B}$ is the magnetic length characterizing the scale
of the wave function, and thus almost all electrons fall into the
LLL in strong magnetic limit. Since the density is spatially
constant, occupied area is exactly determined by fixing the number
of particles. While the area is preserved, positions of particles
can be changed by gauge transformation. Therefore, the electron
state in the strong magnetic field behaves as incompressible fluid.
Although any dynamical degrees of freedom do not exist because we
neglect excitations to higher Landau levels, we should consider
residual degrees of freedom for the fluid, geometrical
configurations of particles, related to area preserving
transformation. Thus Chern-Simons theory which is also non-dynamical
theory captures the feature of the incompressible fluid. An
important property of the incompressible fluid is that it possesses
no dynamical degree of freedom and the residual degree of freedom
comes from geometry of the fluid, which is related to area
preserving diffeomorphism. Indeed one can derive Chern--Simons
action that is the effective theory of the LLL state integrating
over fermion modes \cite{sussk}. We firstly introduce integration
constants of the cyclotron motion describing the residual degrees of
freedom called a guiding center: $X = x + l_0^2\Pi_y, Y = y -
y-l_0^2\Pi_x ,$ where $\vec{\Pi}=\vec{p}+\vec{A}$ is the magnetic
momentum. These operators satisfy the following commutation
relations $[X, Y ] = il_0^2, \quad [\Pi_x,\Pi_y] = - \f{i}{l_0^2}.$
When the magnetic field becomes very strong, contributions of the
magnetic momentum to the guiding center and the canonical momentum
can be neglected so that $\vec{X}\sim x $ and $\vec{p}\sim
-\vec{A}.$ Thus the Lagrangian can be written in terms of the
guiding center coordinates $ {\mathcal
L}=\vec{p}\cdot\vec{x}-{\mathcal H}=\f{B}{2}(X\dot{Y}-\dot{X}Y),
\quad {\mathcal H}=\f{1}{2m}\vec{\Pi}^2 $ and for the $n$-body state
action we have \be S=\f{B}{2}\int
dt\sum_{i=1}^n\varepsilon_{ab}X^a_i\dot{X}^b_i~.\ee In the large $n$
limit, a fluid dynamical description becomes available $\sum_{i=1}^n
\rightarrow \int d^2x\rho(x), \vec{X}_i(t)\rightarrow\vec{X}(x,t),
\vec{X}(x,0)=\vec{x}.$ The initial state is a reference
configuration of the fluid. We will consider fluctuation modes from
the reference state as the residual degree of freedom. The
constraint for the incompressibility is the constant density
condition, $\rho(x)=\rho_e$. Since the density of particles is the
Jacobian of the fluid dynamical field, the constraint can be written
with the Poisson bracket form
\be\rho_e=\rho(x)=\rho_e|\partial\vec{X}/\partial\vec{x}|=\f12\rho_e\varepsilon_{ab}\{X^a,X^b\}.\ee
Adding this Jacobian preservation constraint to action with temporal
gauge field $A_0$ as the Lagrange multiplier, the action is modified
as \be\label{SHall} S=\f{B}{2}\rho_e\int
dtd^2x[\varepsilon_{ab}X^a(\dot{X}^b-\theta\{X^b,A_0\})+2\theta A_0]
,\ee where $\theta=1/2\pi\rho_e$ will become the noncommutative
parameter. The Lagrangian has an exact gauge invariance under area
preserving diffeomorphisms of the $X$ plane. Then, satisfying the
constraint, we can decompose $X^a$ as $X^a = x^a +
\theta^{ab}A_b,\theta^{ab} = \theta\varepsilon^{ab}. $ Here we can
regard gauge fields as the fluctuation mode from the reference
state, and the gauge transformation corresponds to area preserving
transformation of the fluid. Writing the action (\ref{SHall}) in
terms of the gauge fields, we obtain \be S = \f{1}{4\pi\nu}\int
d^3x\varepsilon^{mnl}(\partial_mA_nA_l+\f{\theta}{3}\{A_m,A_n\}A_l).\ee
The constant $1/\nu = 1/(B\theta)=n$ is an integer, which is the
level of the Chern--Simons theory, and $\nu=\rho_e/\rho_0$ is a
filling fraction for the LLL states. Both the odd and even integer
cases describe quantum Hall states, the odd cases corresponding to
fermions and the even to bosons. Furthermore, this action can be
regarded as a leading contribution of the noncommutative
Chern-Simons action \be S=\f{1}{4\pi \nu}\int d^3x
\varepsilon^{mnl}(\partial_m A_n \star A_l-\f{2i}{3}A_m \star
A_n\star A_l),\ee where $\star$-product is the Moyal product defined
as $f(x)\star
g(x)=f(x)\exp(\f{i}{2}\stackrel{\leftarrow}{\partial}_m
\theta^{mn}\stackrel{\rightarrow}{\partial}_n)g(x)$ that exactly
reproduces the quantitative connection between filling fraction
(level in the Chern--Simons description) and statistics required by
Laughlin's theory \cite{sussk}. In this example, we have discussed
the incompressible fluid as the LLL state and its effective theory.

In order to construct non-trivial examples of $n$-algebras,  the
crucial observation that of \cite{takh} where it was noted  that the
Nambu $n$-brackets (\ref{nambr}) in $\mathbb{R}^n$ create a tower of
lower dimensional brackets of order $n-1, n-2, \ldots$ including the
family of Poisson structures on submanifolds which are embedded in
$\mathbb{R}^n.$ Namely, for a fixed $H$ we can define a new bracket
$\{f_1,\ldots,f_{n-1}\}_H=\{H,f_1,\ldots,f_{n-1}\},$ which turns out
to be the Nambu bracket of order $n-1$. Let us consider a smooth
3-manifold ${\cM}_3$ embedded in $\mathbb{R}^4$ through a level-set
Morse function $h(x^1,\ldots,x^4)=c$ with $c\in R$ fixed. Then by
using the fundamental identity (\ref{FINambu}) in $\mathbb{R}^4$ we
can check that the 3-bracket on $\mathbb{R}^4$
\be\label{3-algNP}\{f_1,f_2,f_3\}=\omega^{ijk}(x)\partial_if_1\partial_jf_2\partial_kf_3,
\quad \omega^{ijk}=\epsilon^{ijkl}\partial_l h ,\ee satisfies the
relation \be
\omega^{plm}\partial_p\omega^{ijk}=\omega^{pjk}\partial_p\omega^{ilm}+\omega^{ipk}\partial_p\omega^{jlm}+\omega^{ijp}\partial_p\omega^{klm}
\,.\ee For example if $h$ is a linear function
$h(x^1,\ldots,x^4)=a_i x^i$ then we obtain the constant Nambu
3-algebra $\{x^i,x^j,x^k\}=\varepsilon^{ijkl}a_l.$ If $h$ is a
quadratic function, representing the sphere $S^3\subset
\mathbb{R}^4$: $h=\f12x^ix_i$ then we have the linear Nambu
3-algebra $\{x^i,x^j,x^k\}_{S^3}=\varepsilon^{ijkl}x^l.$ If we use
polar coordinates to project on the surface $e^4=\cos\vartheta_3,$
$e^3=\cos\vartheta_2\sin\vartheta_3,$
$e^2=\sin\vartheta_1\sin\vartheta_2\sin\vartheta_3,$
$e^1=\cos\vartheta_1\sin\vartheta_2\sin\vartheta_3$ then the
3-sphere algebra is
$$\{e^i,e^j,e^k\}_{S^3}=
\f{1}{\sin^2\vartheta_3\sin\vartheta_2}\varepsilon^{pqr}\partial_{\vartheta_p}e^i
\partial_{\vartheta_q}e^j
\partial_{\vartheta_r}e^k=\varepsilon^{ijkl}e^l .$$
By using the Leibniz property, it is possible to write down the
3-algebra on $S^3$ explicitly for a basis of hyperspherical
harmonics the corresponding Nambu $S^3$ 3-algebra
$Y_a=Y_{nlm}(\vartheta_3,\vartheta_2,\vartheta_1)$, $m=-l,\ldots,l$,
$l=0,1,\ldots,n-1,$ \be\{Y_a,Y_b,Y_c\}=f_{abc}^{\ \  d}Y_d~,\ee
where $f_{abc}^{\ \  d}$ can be expressed in terms of 6j symbols of
$SU(2)$ ($SO(4)\sim SU(2)\times SU(2)$).

We observe that, the most general Nambu 3-algebra (\ref{3-algNP})
$\{x^i,x^j,x^k\}_{h}=\epsilon^{ijkl}\partial_lh$ has $h$ as Casimir.
Such restriction of this algebra on the surface $h=c$ gives a
non-degenerate 3-form $\omega^{ijk}$ which satisfies the fundamental
identity. Let us now present two examples of 3-algebras such as
$\mathbb{R}^3$ and $\mathbb{T}^3.$ Obvious that the 3-algebra
$\mathbb{R}^3$ of coordinates is
$\{x^i,x^j,x^k\}=\varepsilon^{ijk}$. For the 3-torus $\mathbb{T}^3$
the algebra for the periodic function basis: $e^n=e^{in\cdot x},$
with $n=(n_1,n_2,n_3)\in \mathbb{Z}^3$ and $x=(x_1,x_2,x_3)\in
(0,2\pi)^3$ is given by $\{e^n,e^m,e^l\}=-in\cdot (m\times
l)e^{n+m+l}.$

In the long history of the study of the Nambu bracket there have
been many attempts to quantize the Nambu mechanics, based on the
deformation theory, path integral formulation and on
Nambu-Heisenberg relation. However, the issue of the quantiza\-tion
is still a difficult child and does not seem to be unique. An
explicit realization of the quantum Nambu bracket in terms of
matrices, as posed in the original paper by Nambu, still seems to be
lacking. As an interesting approach authors \cite{awata} introduced
many-index objects (as particular case three-index objects called
'cubic matrices') to realize the quantum version of the Nambu
bracket. The most mathematically complete quantization scheme for
the Nambu 3-bracket up to now is given in ref. \cite{dito} where an
algebraic topological quantization, the Zariski $\star$ quantization
which is based on factorization of polynomials in several real
variables and variations thereof, has been proposed, but the
algebraic complexity of the scheme seems to hide important physical
and geometrical aspects of the problem. All the other present
proposals are violate, in general, the basic properties of the
3-bracket such as Leibnitz and the Fundamental Identity \cite{takh}.
One of  different approach to quantization is a canonical formalism.
It is based on the Heisenberg commutation relations, which for the
phase space $X=\mathbb{R}^2$ with canonical Poisson bracket look
like the following $[a, a^{\dag}]=I$, where operators $a^\dag,  a$
act in a linear space of quantum states. Being one of a fundamental
principles of quantum mechanics, the Heisenberg commutation
relations have remarkable mathematical properties. In particular,
one has celebrated Stone--von Neumann theorem that all irreducible
representations of the Heisenberg commutation relations are unitary
equivalent. In \cite{Nambu} proposed the following generalization of
the Heisenberg commutation relation \be[\hat{A}_1,\hat{A}_2,
\hat{A}_3]=\hat{A}_1\hat{A}_2\hat{A}_3-\hat{A}_1\hat{A}_3\hat{A}_2+\hat{A}_3\hat{A}_1\hat{A}_2\ee$$-\hat{A}_3\hat{A}_2\hat{A}_1
+\hat{A}_2\hat{A}_3\hat{A}_1-\hat{A}_2\hat{A}_1\hat{A}_3=i\hbar_N
I,$$ where $\hat{A}_1, \hat{A}_2, \hat{A}_3$ are linear operators,
$I$ is a unit and $\hbar_N$ is a constant. Nambu--Heisenberg
relation with $\hbar_n=\sqrt{3}$ admits the following representation
in the Hilbert space ${\mathcal H}_3$
\be\hat{A}_1|\omega>=(\omega+1+\rho)|\omega+1>, \quad
\hat{A}_2|\omega>=(\omega+\rho)|\omega+\rho>,\ee
$$ \hat{A}_3|\omega>=(\omega+\rho^2)|\omega+\rho^2>\,. $$
A direct calculation proves the following result :
$[\hat{A}_1,\hat{A}_2,
\hat{A}_3]|\omega>=\rho^2(1-\rho^2)|1+\rho+\rho^2+\omega>$ with
$1+\rho+\rho^2=0,$ i.e. $\rho=\f{-1+\sqrt{-3}}{2}.$ Then the
Nambu-Heisenberg commutator can have both finite and infinite
dimension linear Hilbert space realiza\-tion   ${\mathcal H}_3$ with
the basis $\{|\omega>\}$ parametrized by  a ring of algebraic
integers $\mathbb{Z}[\rho]$, i.e. \be\omega=m_1+m_2\rho \in
\mathbb{Z}[\rho], \quad m_1,m_2 \in \mathbb{Z}\,.\ee It is
interesting to note that the Nambu--Heisenberg relation suggests the
cubic form of the uncertainty principle $\Delta P\Delta Q\Delta
R\sim\hbar_N$. In \cite{takh} it was also mentioned that the
Nambu--Heisenberg commutation relations for general $n$ admit a
natural representation in the vector space ${\mathcal H}_n$. In
\cite{takh95} has been presented the representation the
Nambu-Heisenberg commutation relation for $n=5$ and $n=7$ which can
be proved directly with the use of a symbolic calculations package
Wolfram Research Mathematica. However, although we are absolutely
certain that there exists a natural representation for any $n$, we
are unable to construct it explicitly.

In this section we presented a geometrical perspective for the
classical and quantum Nambu dynamics in three dimensional phase
space manifolds. The two Hamiltonians are interpreted in following
way: one of the two sets defines the 2- dim phase space geometry
embedded in the 3-dim phase space, while the second one gives the
dynamics of the trajectories on the 2-dim phase space. This view
persists in all higher $n$-dimensions of phase space where  exists
$n-1$ Hamiltonians. Then we choose $n-2$ of them to define a 2-dim
phase space embedded in $n$-dimensions with the $(n-1)$th
Hamiltonians to define the trajectories. This perspective stressed,
in fact, the importance of the $SDiff(M_3)$ group as the all
embracing framework of possible Nambu 3-dim Hamiltonian systems
which, after all, are the flow equations for stationary
incompressible fluids in the manifold. We presented an explicit
const\-ruc\-tions, in the Clebsch--Monge gauge, of the structure
constants of the Nambu 3-algebras for the cases of $\mathbb{R}^3$,
the torus $\mathbb{T}^3$ and the sphere $S^3$ as well as of
quadratic 3-dim manifolds embedded in $\mathbb{R}^4.$  The foliation
of the three dimensional phase space by arbitrary two dimensional
symplectic manifolds, whose quantization is well known either by
operator methods or $\star$ -quantization techniques, motivates the
definition of the quantum 3-bracket (or 3-geometry) as a foliation
of quantum 2-brackets. For this purpose the authors \cite{axe}
define an associative quantization of the algebra
$\{x^i,x^j\}_{H_2}=\varepsilon^{ijk}\partial_k H_2$ promoting the
phase space coordinates $x^i$ at $t=0$ to hermitian operators
$\hat{X}^i$ with commutation relations:
$[\hat{X}^i,\hat{X}^j]=i\hbar\varepsilon^{ijk}P^k(\hat{X})$ having
as a classical limit $\f{1}{i\hbar}[\hat{X}^i,\hat{X}^j]_{\hbar
\rightarrow 0}=\{x^i,x^j\}_{H_2}$. If $H_2$ is a quadratic function
of the canonical phase coordinates there is no ordering problem. For
$H_2$ cubic or higher (non-linear Lie algebra) there is no unique
way to quantize. The quantum 3-commutator should be viewed as the
corresponding quantum volume density element. It is associated, in
our case, with the intersection of quantum (fuzzy) surfaces. It is
hoped that quantum 3-algebras  is a new interesting area of
mathematics in itself, with importance as well for the quantization
of fluid dynamics and more generally for the geometry of 3-d
manifolds (branes) such as our physical space (quantum gravity).

\section{M5 from M2}
When a neutral material put in an external electric field the
electric charges separate from each other and form an electric
dipole. Similar phenomena exists in string theory where there are
extended objects such as $D_p$ branes and higher rank antisymmetric
form fields as charges of these objects. In particular, when a set
of neutral $D$ branes put in an external antisymmetric background
field one observes the ''polarized'' $D$ branes which are expanded
into a higher dimensional world volume theory \cite{myers}. One may
naturally expect such phenomena for membranes in $\cM$-theory.

The eleven dimensional supergravity contains an antisymmetric
three-form field $C_3$ and its magnetic dual six-form field $C_6$.
It is well known that any $p$-brane can covariantly couple to a $p +
1$ form field and also to $p - 1, ...$ form fields due to existence
of world volume antisymmetric gauge fields or Kalb--Ramond fields.
Meyrs in \cite{myers} showed that $p$-brane can also couple to $p +
3, ...$ form fields via the fact that the commutators of transverse
scalar fields in non-Abelian theories are non zero. Then interesting
subject in this crossing pattern is the coupling of supergravity
form fields as background fields with world volume of M2 and M5
branes. By placing a system of $N$ $D_p$ branes in an external
background form field causes that the system would has a vacuum in
which is noncommutative an stable against of the commutative one.
Similarly, in $\cM$-theory, if a collections of membranes would be
in an external $C_6$ form field the system has a vacuum in which
membranes are polarized due to field strength effect and are formed
into a fuzzy $S^3$ sphere. This can be interpreted as the formation
of spherical branes ended on five-brane. Thus, a single M5-brane may
contain multiple M2-branes and is therefore a promising starting
point for a construction of the BLG model. Another indication of
this is that the Nambu-bracket realization of the BLG theory
introduces some 'internal' Riemannian 3-manifold $\cN_3,$ so that
the total space dimension is $2+3=5.$ In this realization, the BLG
model is essentially an exotic gauge theory for the group
$SDiff(S^3)$ of volume-preserving diffeomorphisms of the 3-sphere.

This is not the first occasion on which exotic gauge theories based
on volume-preserving diffeomorphisms have appeared \cite{town88}.
One can use various methods to study M5 brane theory by using M2
theory  but the authors of \cite{ho} described an interesting
approach to achieve this goal using the Nambu 3-brackets. In fact,
by considering an 3 dimensional internal space in the world volume
of M2 brane, they were able to find a six dimensional theory which
has some desired properties of an M5 brane. For example, they found
the action of a self dual two-form gauge field living on the world
volume of M5 brane.

For the construction of M5-brane, the authors \cite{ho} introduce an
''internal'' three-manifold $\cN$ and use the Nambu 3-bracket
$$\{f,g,h\}=P^{\mu\nu\lambda}(y)\partial_\mu f\partial_\nu g\partial_\lambda h$$
on $\cN$ as a realization of three-algebra. Here $y^\mu$
($\mu=1,2,3$) is the local coordinates on $\cN$. One of the most
important properties of the Nambu 3-bracket is that it satisfies the
analog of the fundamental identity for arbitrary functions $f_i$ on
$\cN$,
\be\label{fiNP}\{f_1,f_2\{f_3,f_4,f_5\}\}=\ee$$\{\{f_1,f_2,f_3\},f_4,f_5\}+\{f_3,\{f_1,f_2,f_4\},f_5\}+\{f_3,f_4\{f_1,f_2,f_3\}\}~.$$
This gives a very severe constraint on the coefficient
$P^{\mu\nu\lambda}(y).$ Actually it is known that by the suitable
choice of the local coordinates, it can be reduced to the Jacobian
\be\label{3dNP}\{f,g,h\}=\varepsilon^{\mu\nu\lambda}\f{\partial
f}{\partial y^\mu}\f{\partial g}{\partial y^\nu}\f{\partial
h}{\partial y^\lambda}.\ee This property is referred to as the
'decomposability' in the literature. If we choose the basis of
functions on $\cN$ as $\chi(y)$ ($a = 1, 2, \ldots$) and write the
Nambu-Poisson bracket as a Lie 3-algebra,
\be\{\chi^a,\chi^b,\chi^c\}=\varepsilon^{\mu\nu\lambda}\partial_\mu
\chi^a\partial_\nu \chi^b\partial_\lambda \chi^c=f^{abc}_{\ \ \
d}\chi^d(y)~.\ee Eq. (\ref{fiNP}) implies that the structure
constant $f^{abc}_{\ \ \  d}$ here satisfies the fundamental
identity. The integration $<f,g>=\int_{\cN}d^3y f(y)g(y)$ over the
$y$-space can be used to define the invariant metric
$h^{ab}=(\chi^a, \chi^b).$ Except for the trivial case (${\cN} =
\mathbb{R}^3$), we have to cover $\cN$ by local patches and the
coordinates $y^\mu$ are the local coordinates on each patch. If we
need to go to the different patch where the local coordinates are
$y^{'\mu}$, the coordinate transformation between $y$ and $y^{'}$
(say $y^{'\mu} = f^\mu (y)$) should keep the Nambu 3-bracket
(\ref{3dNP}). It implies that $\{f^1,f^2,f^3\}=1$. Namely $f^\mu
(y)$ should be the volume-preserving diffeomorphisms. As we will
see, the gauge symmetry of the BLG model for this choice of the
Filippov-Nambu 3-algebra is the volume-preserving diffeomorphisms of
$\cN$ which is very natural in this set-up.

Now, we will show that the BLG model with a Filippov-Nambu structure
on a 3-dimensional manifold contains the low energy degrees of
freedom on an M5-brane. Before going on, let us count the number of
degrees of freedom in the bosonic and fermionic sectors in our
model. The fermion is a Majorana spinor in 10+1 dimensions with a
chirality condition, and thus it has 16 real fermionic components,
equivalent to 8 bosonic degrees of freedom. For a 5-brane there are
5 transverse directions corresponding to 5 scalars $X^i$. For an
ordinary 2-form gauge field in 6D, there are 6 propagating modes.
But since we do not have the usual kinetic term for $A_m$, but
rather a Chern---Simons term, there are only 3 propagating modes.
The low energy effective theory of an M5-brane contains the same
number of bosonic and fermionic degrees of freedom. But a salient
feature of the M5-brane is that the 2-form gauge field is self-dual.
Hence our major challenge is to show that the gauge field of the BLG
model is equivalent to a self-dual 2-form gauge field.

A comment on the notation: we will use $I,J,K$ to label the
transverse directions to the membrane worldvolume $\cM$. We
decompose this eight dimensional space as a direct product of $\cN$
and remaining 5 dimensional space. We use $\mu, \nu, \lambda$ to
label $\cN$ and $i, j, k$ to label the transverse directions of the
M5-brane. By combining the basis of $C({\cN})$, we can treat
$X^I_a(x)$ and  $\Psi_a(x)$ as six-dimensional local fields \be
X^I(x,y)=\sum_a X^I_a(x)\chi^a(y), \quad \Psi(x,y)=\sum_a
\Psi_a(x)\chi^a(y)~.\ee Similarly, the gauge field $A^{ab}_m$ can be
regarded as a bi-local field: \be
A_m(x,y,y')=A^{ab}_m(x)\chi^a(y)\chi^b(y')~.\ee The existence of
such a bi-local field does not mean that the theory is non-local.
Let us expand it with respect to $\Delta y^\mu=y^{'\mu}-y^\mu$ as
$$A_m(x,y,y')=a_m(x,y)+b_{m\mu}(x,y)\Delta y^\mu +\f12 c_{m\mu\nu}(x,y)\Delta y^\mu\Delta y^\nu+\ldots$$
Because $A_{m}^{ab}$ always appears in the action in the form
$f^{bcd}_{\ \ \  a}A_{m\; bc}$, the field $A_m(y,y')$ is highly
redundant, and only the component
$b_{m\mu}(x,y)=\f{\partial}{\partial y^{'
\mu}}A_{m}(x,y,y')|_{y'=y}$ contributes to the action. For example,
the covariant derivative of BLG model is rewritten for our case as,
\be\label{cov-b}\cD_m X^I(x,y)=(\partial_m X^I_a(x)-gf^{bcd}_{\ \ \
a}A_{m \; bc}X^I_d(x))\chi^a(y)\ee
$$=\partial_m X^I(x,y)-g\varepsilon^{\mu\nu\rho}\f{\partial^2 A_m(x,y,y')}{\partial y^\mu\partial y^{'\nu}}|_{y=y'}\f{\partial X^I(x,y)}{\partial y^\rho}$$
$$=
\partial_m X^I(x,y)-g\varepsilon^{\mu\nu\rho}\partial_\mu b_{m\nu}(x,y)\partial_\rho X^I(x,y)=\partial_m X^I-g\{b_{m\nu}, y^{\nu},
X^I\}~.$$ The covariant derivative for the fermion field is
similarly, \be\cD_m\Psi(x,y)=\partial_m\Psi(x,y)-g\{b_{m\nu},
y^{\nu}, \Psi\}~.\ee

In \cite{ho}, this theory written in terms of fields on six
dimensions is identified with the theory describing a single
M5-brane. However we still have SO(8) global  symmetry, which is
different from the SO(5) symmetry expected in the M5-brane theory.
Then to interpret the six-dimensional theory, we must take a partial
static gauge for three among six world-volume coordinates. As we
mentioned above, however, we do not have full diffeomorphisms in the
$y^\mu$ space. The action is invariant only under volume-preserving
diffeomorphisms. This implies that we cannot comple\-tely fix the
fields $X^\mu$, and there are remaining physical degrees of freedom.
For this reason, we should loosen the static gauge condition as \be
X^\mu(x,y)=\f{1}{g}y^\mu +b^\mu(x,y), \quad
b_{\mu\nu}=\f12\varepsilon_{\mu\nu\rho}b^\rho~.\ee As was shown in
\cite{ho}, the tensor field $b_{\mu\nu}$ is identified with a part
of the 2-form gauge field on a M5-brane. The gauge transformations
of the scalar fields $X^I$ and fermion fields   are given by
\be\label{diff} \delta_\Lambda X^I(x,y)=g \Lambda_{ab}(x)f^{abc}_{\
\ \ d}X^I_c(x)\chi^d(y)=g\Lambda_{ab}(x)\{\chi^a,\chi^b, X^I\}\ee$$=
g(\delta_\Lambda y^\mu)\partial_\mu X^I(x,y), \quad \delta_\Lambda
\Psi(x,y)=g(\delta_\Lambda y^\mu)\partial_\mu \Psi(x,y),$$ where we
used $f^{abc}_{\ \ \ d}=(\{\chi^a, \chi^b,\chi^c\},\chi_d)$, $
\sum_a\chi^a(y)\chi_a(y')=\delta(y-y')$ and $\delta_\Lambda y^\rho$
is defined as
$$\delta_\Lambda y^\rho=\varepsilon^{\rho\mu\nu}\partial_\mu\Lambda_\nu(x,y),
\Lambda_\mu(x,y)=\partial'_\mu\tilde\Lambda(x,y,y')|_{y=y'},
\tilde\Lambda(x,y,y')=\Lambda_{ab}(x)\chi^a(y)\chi^b(y').$$ The
transformation (\ref{diff}) may be regarded as the infinitesimal
reparametrization $y^{'\mu}=y^\mu-g\delta y^\mu$. Since
$\partial_\mu\delta y^\mu=0$, it represents the volume-preserving
diffeomorp\-hisms. As the symmetry is local on $\cM$, the gauge
parameter is an arbitrary function of $x$. So what we have obtained
is a gauge theory on $\cM$ whose gauge group is the
volume-preserving diffeomorphisms of $\cN$ which preserves the
volume form $\omega=dy^1\wedge dy^2\wedge dy^3$.

The following combination defines the 'covariant' derivative along
the fiber direction: \be\cD_\mu \Phi\equiv \f12
g^2\varepsilon_{\mu\nu\rho}\{X^\nu,X^\rho,
\Phi\}\ee$$=\partial_\mu\Phi+g(\partial_\lambda
b^\lambda\partial_\mu-
\partial_\mu b^\lambda\partial_\lambda )\Phi+\f12 g^2\varepsilon_{\mu\nu\rho}\{b^\mu,b^\nu, \Phi\}.$$
Together with (\ref{cov-b}) we have a set of covariant derivatives
on M5 world-volume. Just like the case of ordinary gauge theories,
the field strength of the tensor field $\mathcal H$ arises in the
commutator of the covariant derivatives defined above:
\be\label{kom} [\cD_\mu,
\cD_\nu]\Phi=g^2\varepsilon_{\mu\nu\rho}\{{\mathcal H}_{123},
X^\rho, \Phi\}, \quad [\cD_m, \cD_\mu]\Phi =g^2\{{\mathcal
H}_{m\rho\mu}, X^\rho, \Phi\},\ee
$$[\cD_m, \cD_n]\Phi=-\f{g}{V}\varepsilon_{mnl}\cD_p\tilde{\mathcal H}^{p\;l\mu}\cD_\mu\Phi,$$
where $V$ is the 'induced volume' $V = g^3\{X^1 ,X^2 ,X^3 \}$, and
$\tilde{\mathcal H}$ is a dual  field strength. Equation
(\ref{kom}), in which $\Phi$ is taken to be $X^\mu$ is nothing but
the Bianchi identity $\cD_l \tilde{\mathcal
H}^{lmn}+\cD_\mu\tilde{\mathcal H}^{\mu mn}\equiv 0, $ where
$\tilde{\mathcal H}^{lmn}$ and $\tilde{\mathcal H}^{\mu mn}$ are
Hodge dual of ${\mathcal H}_{\mu\nu\lambda}$ and ${\mathcal
H}_{\mu\nu m}$. Now we rewrite the various parts of the BLG action
in terms of the six dimensional fields and their covariant
derivatives \be S_X+S_{pot}=\int d^3x<-\f12(\cD_m
X^i)^2-\f12(\cD_\mu X^i)^2-\f14{\mathcal
H}^2_{m\mu\nu}-\f{1}{12}{\mathcal H}^2_{\mu\nu\rho}-\f{1}{2g^2}\ee
$$-\f{g^4}{4}\{X^\mu,X^i,X^j\}^2-\f{g^4}{12}\{X^i,X^j,X^k\}^2>$$
\be S_\Psi +S_{int}=\int d^3x
<\f{i}{2}\bar\Psi\Gamma^m\cD_m\Psi+\f{i}{2}\bar\Psi\Gamma^\rho\Gamma_{123}\cD_\rho\Psi\ee
$$ +i\f{g^2}{2}\bar\Psi\Gamma_{\mu i}\{X^\mu,X^i,\Psi\}
+i\f{g^2}{4}\bar\Psi\Gamma_{ij}\{X^i,X^j, \Psi\}> ,$$ where
$<f,g>=\int d^3y f g.$ The Chern--Simons term cannot be rewritten in
manifestly gauge-covariant form \be S_{CS}=\int d^3x
\varepsilon^{mnp}<-\f12\varepsilon^{\mu\nu\lambda}\partial_\mu
b_{m\nu}\partial_n b_{p\lambda}+\f{g}{6}\varepsilon^{\mu\nu\lambda}
\partial_\mu b_{n\nu}\varepsilon^{\rho\sigma\tau}\partial_\sigma b_{p\rho}(\partial_\lambda b_{m\tau}-\partial_\tau b_{m\lambda})>~.\ee
However, the equation of motion which is derived from these actions
turns out to be manifestly gauge-covariant. This was confirmed
recently in \cite{sams} where it was shown that solving the field
equations associated with $b_{m\mu}$ and $b_{\mu\nu}$ is tantamount
to imposing the Hodge self-duality condition on the non-linear field
strength. This allowed the authors to rewrite the gauge field
Lagrangian in a gauge covariant form as \be S=-\int d^3xd^3y
\{\f18{\mathcal H}_{m\mu\nu}{\mathcal
H}^{m\mu\nu}+\f{1}{12}{\mathcal H}_{\mu\nu\rho}{\mathcal
H}^{\mu\nu\rho}-
\f{1}{144}\varepsilon^{mnl}\varepsilon^{\mu\nu\rho}{\mathcal
H}_{mnl}{\mathcal H}_{\mu\nu\rho}\ee
$$-\f{1}{12 g}\varepsilon^{mnl}{\mathcal H}_{mnl}\}.$$
The last term in this expression can be interpreted as a coupling of
the M5-brane to the constant background $C_3$ field which has
non-zero components $C_{mnl}=\f{1}{g}\varepsilon_{mnl}.$ Following
\cite{sams}, it is possible to rewrite this as $\f12\int {\mathcal
H}_3\wedge C_3.$ This action possesses full volume preserving
diffeomorphism invariance. However the Lorentz symmetry is broken by
the presence of the three-form field.

Next we can rewrite the supersymmetry transformations in terms of
the six-dimensional covariant derivatives and field strength:
$$\delta X^i=i\bar\epsilon \Gamma^i\Psi, \quad \delta b_{\mu\nu}=-i\bar\epsilon \Gamma_{\mu\nu}\Psi, \quad \delta b_{m\nu}
=-iV(\bar\epsilon\Gamma_m\Gamma_\nu \Psi)+ig(\bar\epsilon
\Gamma_m\Gamma_i\Gamma_{123}\Psi)\partial_\nu X^i,$$
$$\delta\Psi=\cD_m X^i\Gamma^m\Gamma^i\epsilon +\cD_\mu X^i \Gamma^\mu\Gamma^i\epsilon
-\f12{\mathcal H}_{m\nu\rho}\Gamma^m\Gamma^{\nu\rho}\epsilon
-(\f{1}{g}+{\mathcal H}_{123}) \Gamma_{123}\epsilon$$$$
-\f{g^2}{2}\{X^\mu,X^i,X^j\}\Gamma^\mu\Gamma^{ij}\epsilon
+\f{g^2}{6}\{X^i,X^j,X^k\}\Gamma^{ijk}\Gamma^{123}\epsilon.$$ A
peculiar property of this supersymmetric transformation is that the
perturbative vacuum (the configuration with all fields vanishing) is
not invariant under this transformation due to the term in
$\delta\Psi$ proportional to $1/g$. We can naturally interpret this
term as a contribution of the background C-field. In the M5- brane
action coupled to background fields, the self-dual field strength is
defined by $H = db + C$. The inclusion of C-field in the field
strength is required by the invariance of the action under C-field
gauge transformations. The shift of the field strength ${\mathcal
H}_{123}$ by $1/g$ in the action as well as in the supersymmetric
transformation suggests that the relation $C \sim g^{-1}$ between
the Nambu structure and the $C$-field background. In fact, M5-brane
in a constant $C$-field background is still 1/2 BPS. The effect of
the C-field is changing which half of 32 supersymmetry remain
unbroken. We can find this phenomenon in our six-dimensional theory.
In addition to 16 supersymmetries we described above, the theory has
16 non-linear fermionic symmetries $\delta^{nl}$, which shift the
fermion by a constant spinor $\delta^{nl}\Psi=\xi~.$ The action is
invariant under this transformation because constant functions in
$y^\mu$ space are in the center of the 3-algebra. The perturbative
vacuum is invariant under the combination of two fermionic
symmetries $\delta_\epsilon -\f{1}{g}\delta^{nl}.$ In the weak
coupling limit $g \rightarrow 0$, the transformation laws for this
combined symmetry agree with those of an $N = (2, 0)$ tensor
multiplet:
$$\delta X^i=i\bar\epsilon \Gamma^i\Psi, \quad
\delta
\Psi=\partial_{\underline{\mu}}X^i\Gamma^{\underline{\mu}}\Gamma^i\epsilon
-\f{1}{12}H_{\underline{\mu\nu\rho}}\Gamma^{\underline{\mu\nu\rho}}\epsilon
\quad \delta
b_{\underline{\mu}\underline{\nu}}=-i\bar\epsilon\Gamma_{\underline{\mu}\underline{\nu}}\Psi~.$$

The gauge symmetry of the M5 world-volume theory is the
volume-preserving diffeomorphisms on $\cN$. The transformation law
for both $X^i$ and  $\Psi$    are given in the same form
(\ref{diff}) where the volume-preserving coordinate transformation
is parametrized by three arbitrary functions $\Lambda_\mu$. While
$b_{m\mu}$ and $b_{\mu\nu}$ are viewed as the gauge potentials for
the gauge symmetry of coordinate transformations preserving a given
Nambu structure, $B_m^\mu$ and $b^\mu$ should be viewed as two types
of deformation parameters of the Nambu structure of the M5-brane
world-volume. We have $b^\mu$ specifying the change of the Nambu
structure due to a change of coordinates $\delta y^\mu$ in $\cN$ (so
that the volume form is changed), and $B_m^\mu$ specifying the
change due to a mixing of the two classes of coordinates $x^m$ and
$y^\mu$. The gauge symmetry corresponds to redundant descriptions of
deformations of the Nambu structure. Then the M5-brane theory with a
self-dual gauge field can thus be interpreted as a dynamical theory
of the Nambu structure.

\section{Reformulation of Dirac--Nambu--Goto action by Nambu bracket}
In the late 80's in \cite{gsw} has been remarked that "Eleven
dimensional supergravity remains enigma". In a papers \cite{town88}
(and references therein) it was suggested that, just as
10-dimensional supergravity is related to superstring theory, so
11-dimensional supergravity may be related to supermembrane theory.
In support of this connection it was built an 11-dimensional
supermembrane action and shown that the preser\-va\-tion of local
symmetries of this action in an 11-dimensional background requires
that the background satisfy certain constraints, which are
equivalent to the equations of motion of 11-dimensional
supergravity. Furthermore, it was argued that the spectrum of the
11-dimensional supermembrane contains the massless states of 11$d$
supergravity. Then one can hope that a supermembrane theory will
provide a quantum consistent extension of 11-dimensional
supergravity just as superstring theories are thought to provide a
quantum consistent extension of 10-dimensional supergravity
theories. Already classically the possibilities for super $p$-brane
actions are severely limited. It is well-known that the
Green--Schwarz superstring action exists for $d=3,4,6,10$ and one
can similarly show that the supermembrane action exists for
$d=4,5,7,11.$ Then we might expect quantum considerations to impose
yet further restrictions. Indeed we know that only the 10$d$
superstring action is quantum consistent (i.e. free from anomalies).
This might lead one to suspect that the only quantum consistent
super $p$-brane is 11$d$ supermembrane. However, despite of its
elegant geometric significance, Nambu-Goto action for $p>1$ is
difficult to quantize because its highly nonlinear structure. In the
remarkable  papers \cite{park08}, \cite{park} the authors
constructed an action whose characteristic features are the
appearance of gauge covariant derivatives and the Nambu bracket
squared potential. After some gauge fixing, the action can be
identified as a lower dimensional gauge theory action based on
Filippov-Lie algebra. Further, in order to emphasize significance of
the results given in \cite{park08}, \cite{park} we will literally
cite the materials from these papers.

With an embedding of $(p + 1)$-dimensional worldvolume coordinates
into D-dimensional target spacetime, $X(\xi): \xi^m \rightarrow
X^M$, where $m=0,1, \ldots,p$ and $M=0,1, \ldots, D-1$, the
Dirac--Nambu--Goto (DNG) action for a $p$-brane reads \cite{polch}
\be\label{actDNG} S_{DNG}=-T\int d^{p+1}\xi\sqrt{-\det(\partial_m
X^M\partial_n X_M)}~,\ee where $T$ is the membrane tension. Let us
decompose, formally, the $p$-brane world\-vo\-lume coordinates into
two parts: $\{\xi^m \}=\{\sigma^\mu,\varsigma^i\}$, where $\mu=0,1,
\ldots, d-1$ and $i=1, \ldots, \hat{d}$. The decomposition is a
priori arbitrary for any non-negative integers $d, \hat{d}$. One
natural application of the splitting will be the case where p-brane
is extended over two topologically different spaces, e.g. compact
and non-compact spaces. With the decomposition above, a square root
free reformulation of the DNG action was achieved in:
\be\label{N-G}S=\int d^d\sigma \mbox{Tr}(\sqrt{-h}{\mathcal L}),
\quad\quad \mbox{Tr}:=\int d^{\hat{d}}\varsigma, \quad {\mathcal
L}=-h^{\mu\nu}\cD_\mu X^m\cD_\nu X_M\ee
$$-\f{1}{4\hat{d}!}
e^{d-1}\{X^{M_1},X^{M_2},\ldots,X^{M_{\hat{d}}}\}
\{X_{M_1},X_{M_2},\ldots,X_{M_{\hat{d}}}\}+(d-1)e\,,
$$
where the action contains three kinds of auxiliary fields: scalar
$e$, $d$-dimensional metric $h_{\mu\nu}$ and a gauge connection
$A^i_\mu$ which defines the ‘covariant derivative’: $\cD_\mu
X^M:=\partial_\mu X^M-A^i_\mu\partial_iX^M$. The classical equation
of motion that follows from (\ref{actDNG}) may equivalently be
obtained from the action (\ref{N-G}). Integrating out all the
auxiliary fields, using their on-shell values, the action reduces to
the DNG action, $S\equiv S_{DNG}$, and hence the classical
equivalence. The novelty \cite{park08} of the above reformulation
was the appearance of the gauge interaction and the Nambu bracket
squared potential. The latter basically stems from an identity
rewriting the deter\-mi\-nant as the Nambu bracket squared
\cite{park08}:
\be\label{ident-det}\det(\partial_iX^M\partial_jX_M)=\f{1}{\hat{d}!}
e^{d-1}\{X^{M_1},X^{M_2},\ldots,X^{M_{\hat{d}}}\}
\{X_{M_1},X_{M_2},\ldots,X_{M_{\hat{d}}}\}.\ee A physical picture
after the reformulation can be  described as a single brane as a
conden\-sa\-tion of multiple lower-dimensional branes, i.e. a
$p$-brane by $(d-1)$-branes. Obviously, the choice of $\hat{d} = 0$
and $d = p + 1$ corresponds to the well-known "Polyakov" action,
which was actually first conceived by the authors \cite{brink}. On
the other hand, with a gauge fixing for $e$ to be constant, the
other extreme choice of $d = 0$, $\hat{d} = p + 1$ leads to the
Schild action. Furthermore, the association of the digits, $2$ and
$3$ to string and ${\cM}$-theory becomes manifest within this
reformulation \cite{park08}, \cite{park}. For example, the fact that
the codimension of $D$-branes is 2 suggests to choose $ \hat{d} =
2$, which leads to the two-algebra as in the Yang--Mills theory.
Likely the choice of $p = 5$, $d = 3$, $\hat{d} = 3$ suggests that
the BLG model with an infinite dimensional gauge group describes a
M5-brane as a condensation of multiple M2-branes. Here we presented
a generalization of the Polyakov method, a novel scheme
\cite{park08} to take off the square root of DNG action for a
$p$-brane. While the square root free Polyakov action is a
$(p+1)$-dim field theory, the resulting action (\ref{N-G}) lives in
an arbitrary lower dimensional $d$ which is smaller than $p+1$. Such
a reformulation shows a the general phenomenon that non-Abelian
structure of lower dimensional gauge theories can capture the
description of higher dimensional objects. It suggests that a single
$p$-brane can be described via different but equivalent actions,
either $(p + 1)$-dimensional Polyakov action or various lower
dimensional gauge theories with the Nambu bracket interactions of
different degrees. This implies the existence of a web of duality
relations among large classes of gauge theories. In particular, a
theory with the Yang--Mills interaction, i.e. Poisson bracket of
$\hat{d} = 2$, is equivalent to lower dimensional theories based on
Nambu bracket structure.

The reformulation of the DNG action (\ref{N-G}) is purely bosonic.
In order to establish a firm connection to string/$\cM$-theory one
needs to supersymmetrize them. The requirement of supersymmetry may
give rise to a constraint on the a priori arbitrary decomposition,
$p + 1 = d + \hat{d}$. Our main interest is to supersymmetrize the
action (\ref{N-G}). For $d = 1$ case, supersymmetric actions are
ready to be read-off from an earlier work \cite{berg}. These authors
listed light-cone gauge fixed supersymmetric actions for various
$p$-branes in diverse spacetime dimensions. As usual, the Fierz
identity required for the supersymmetry invariance, restricts the
possible values of $p$ and the spacetime dimension $D$: \be p=1,
\quad D=3,4,6,10 ;\quad \quad p=2, \quad D=4,5,7,11;\ee
$$p=3, \quad D=6,8 ;\quad \quad p=4, \quad D=9;\quad \quad p=5, \quad D=10. $$

In the string theory, the lightcone gauge
$X^+=\f{1}{\sqrt{2}}(X^0+X^{D-1})=\tau$ is convenient for
quantization because it allows the elimination of all unphysical
degrees of freedom and unitarity is guaranteed. Of course, one loses
manifest Lorentz invariance and one must be careful to check that it
is not destroyed by quantization. In membrane theory, however, the
lightcone gauge does not eliminate all unphysical degrees of
freedom. For membranes, however, only $(D - d)$ variables are
physical. Thus the lightcone gauge must leave a residual gauge
invariance \cite{town88}. Utilizing the identity (\ref{ident-det}),
in terms of the Nambu $p$-bracket, their light-cone gauge fixed
supersym\-met\-ric $p$-brane actions can be reexpressed in a compact
form \cite{park}: \be\label{ligcon}{\mathcal L}_{L.C.}=\f12(\cD_\tau
X^I)^2-\f{1}{2p!}\{X^{I_1},\ldots,
X^{I_p}\}^2+\f{i}{2}\bar\Psi\cD_\tau\Psi\ee$$+\f{1}{2(p-1)!}\bar\Psi\Gamma^{I_1\ldots
I_{p-1}}\{X_{I_1},\ldots, X_{I_{p-1}}\Psi\},$$ where
$\cD_\tau=\partial_\tau +u^a(\sigma,\tau)\partial_a$ is a 'covariant
time derivative' with 'gauge field' $u^a$ satisfying $\partial_a
u^a=0.$ For a membrane ($p=2$) of spherical topology, the solution
of this constraint is $u^a =\varepsilon^{ab}\partial_b \omega.$
Remarkably that for correspondence $X^I\rightarrow A^I, \omega
\rightarrow A_0$, this looks like a (D - 1) dimensional
supersymmetric Yang--Mills theory  dimensionally reduced to one time
dimension with infinite dimensional gauge group. This group is, in
fact, the subgroup of the worldvolume diffeomorphisms group that
preserves the Nambu bracket $\{f,g\}=\varepsilon^{ab}\partial_a
f\partial_b g$ and is known as the group of area-preserving
diffeomorphisms. The nature of this infinite-dimensional gauge group
depends criti\-cal\-ly on the topology of the membrane. As example,
for spherical topology it was shown to be $SU(\infty)$ by Hoppe
\cite{hop}. This has an important application in regularization of
membrane theories by replace the gauge theory of $SDiff(S^2)$ by
gauge theory of $SU(N)$. Hoppe has studied the canonical
quantization of a relativistic spherical membrane in the light cone
gauge. He find that the classical $SU(N)$ Yang--Mills theories, in
the large $N$ limit \be\lim_{N\rightarrow\infty}
N[A_\mu,A_\nu]=\{A_\mu,A_\nu\}~,\ee can be described as a new type
of gauge principle. The gauge potentials become c-number functions
\be A_\mu(x,\vartheta,\varphi)=\sum_{l=1}^\infty\sum_{m=-l}^l
A^{lm}_\mu(x)Y_{lm}(\vartheta,\varphi)~,\ee of two additional
coordinates, which parametrize the surface of an internal sphere at
every space-time point. The new gauge transformations \be\delta
A_\mu(x,\vartheta,\varphi) =\partial_\mu
v(x,\vartheta,\varphi)+\{A_\mu, v\}~,\ee
 where the Poisson bracket
of two functions is defined as $\{f,g\}= \f{\partial f}{\partial
\cos\vartheta}\f{\partial g}{\partial \varphi}-\f{\partial
f}{\partial \varphi}\f{\partial g}{\partial \cos\vartheta}~,$ is
isomorphic to the infinite dimensional Lie algebra of  area
preserving (or symplectic) diffeomorphisms of the sphere
$SDiff(S^2)$ which is the symmetry of the membrane after gauge
fixing. In the case  $p$-brane $u^a$ can be written in terms of
functions $A_k$, ($k=1,\ldots, p-1$) as \be
u^a=\varepsilon^{a_1\ldots a_{p-1}a}\f{\partial A_1}{\partial
\sigma^{a_1}} \ldots\f{\partial A_{p-1}}{\partial
\sigma^{a_{p-1}}}~,\ee and then the covariant time-derivative can be
written in the form \be \cD_\tau X^i=\f{\partial X^i}{\partial \tau}
+\{A_1,\ldots,A_{p-1}, X^i\}~.\ee As a result, the action
(\ref{ligcon}) is invariant under the $p$-dimensional volume
preserving diffeo\-mor\-phisms: \be \delta X^i=\lambda^a\partial_a
X^i, \quad \delta
u^a=-\partial_\tau\lambda^a-u^b\partial_b\lambda^a+\lambda^b\partial_bu^a~.\ee
In this case $\lambda^a$ is written in terms of functions
$\Lambda_k$, ($k=1,\ldots,p-1$) in the same way as for $u^a$ and the
transformation laws of $p$-dimensional volume preserving
diffeomorphisms are rewritten as \be\delta
X^i=\{\Lambda_1,\ldots,\Lambda_{p-1}, X^i\}, \ee$$ \delta
u^a=-\partial_\tau\lambda^a-\{A_1,\ldots,A_{p-1},\lambda^a\}+
\{\Lambda_1,\ldots,\Lambda_{p-1},u^a\}.$$

In the  paper \cite{park}, authors consider an alternative choice of
$d = 0.$ In particular, they focus on a supermembrane propagating in
eleven-dimensional flat spacetime and proposed to following action
for the three-algebra description of a supermembrane in eleven
dimensions: \be\label{actNG} S_{M2}=\int d^3 \xi ({\mathcal
L}_\omega +{\mathcal L}_{WZ}),\ee
$${\mathcal L}_\omega=\f{1}{12}\omega^{-1}<E^M,E^N,E^P><E_M,E_N,E_P>-\f12 \omega,$$
$${\mathcal L}_{WZ}=-\f{i}{2}\varepsilon^{ijk}\bar\theta\Gamma_{MN}\partial_i\theta (E^M_j\partial_k X^N-
\f13\bar\theta \Gamma^M\partial_j\theta
\bar\theta\Gamma^N\partial_k\theta),$$ which contains
eleven-dimensional target spacetime coordinates $X^M$, a Majorana
spinor $\theta$ and a scalar density field $\omega$. The former two
are dynamical while the last one is auxiliary. With the
supersymmetry invariant pull-back
$$E^M_i=\partial_i X^M-i\bar\theta\Gamma^M\partial_i\theta,$$
the authors \cite{park} set
$$<E^M,E^N,E^P>=\varepsilon^{ijk}E_i^M E_j^NE_k^P,$$
which has the following expansion in terms of the Nambu-bracket
$$<E^L,E^M,E^N>=\{X^L,X^M,X^N\}-3i\bar\theta\Gamma^{[L}\{X^M,X^{N]},\theta\}
$$$$+3\bar\theta\{\Gamma^{[L}\theta,X^M,\bar\theta\Gamma^{N]}\}\theta-i\bar\theta_\alpha\bar\theta_\beta\bar\theta_\gamma
\{(\Gamma^{[L}\theta)^\alpha,(\Gamma^{M}\theta)^\beta,(\Gamma^{N]}\theta)^\gamma\}~.$$
Similarly, the Wess--Zumino part of the action can be also
reexpressed in terms of the Nambu-bracket: \be{\mathcal
L}_{WZ}=-\f{i}{2}\bar\theta\Gamma_{MN}\{X^M,X^N,\theta\}+\f12\bar\theta_\alpha\bar\theta_\beta\{(\Gamma_{MN}\theta)^\alpha,(\Gamma^M\theta)^\beta,
X^N\}\ee
$$-\f{i}{6}\bar\theta_\alpha\bar\theta_\beta\bar\theta_\gamma\{(\Gamma_{MN}\theta)^\alpha,(\Gamma^{M}\theta)^\beta(\Gamma^{N}\theta)^\gamma\}~.$$
Thus, all the derivatives appear only through the Nambu
three-brackets. Let us now introduce a shorthand notation for the
induced metric: $g_{ij}=E^M_iE_{M j}$ and denote its determinant by
$g = \mbox{det}(g_{ij})$ as usual. All the equations of motion are
then summarized by: \be\omega-\sqrt{-g}=0, \quad
g^{ij}E^M_i\Gamma_M(1-\Gamma)\partial_j \;\theta=0, \quad
\partial_i(\sqrt{-g}g^{ij}E^M_j)-i\varepsilon^{ijk}\partial_i\bar\theta\Gamma^M_{\
\ N}
\partial_j\theta\Pi^N_k=0~.\ee
From an identity
$\f16<E^M,E^N,E^P><E_M,E_N,E_P>=\mbox{det}(E^M_iE_{M j})$
integrating over the auxiliary scalar assuming the on-shell value
$\omega=\sqrt{-g}$, this proposed action (\ref{actNG}) \cite{park}
reduces to the well-known supersymmetric DNG action for M2-brane
\cite{town88}: \be S_{M2}=\int d^3\xi[-\sqrt{-\mbox{det}(E^M_iE_{M
j})}-\f{i}{2}\varepsilon^{ijk}\bar\theta\Gamma_{MN}\partial_i\theta(E^M_j\partial_kX^N-
\f13\bar\theta \Gamma^M\partial_j\theta
\bar\theta\Gamma^N\partial_k\theta)]~.\ee The action (\ref{actNG})
is invariant under the following transformations:

a) Target-spacetime supersymmetry:
\be\delta_\epsilon\theta=\epsilon, \quad \delta_\epsilon
X^M=-i\bar\theta\Gamma^M\epsilon, \quad \delta_\epsilon
\omega=0~;\ee

b) Local 32-component fermionic symmetry:
\be\delta_\zeta\theta=(1+(\omega/\sqrt{-g}) \Gamma)\zeta, \quad
\delta_\zeta X^M= i\bar\theta\Gamma^M\delta_\zeta\theta, \quad
\delta_\zeta\omega=4i\omega(g^{-1})^{ij}
E^M_i\partial_j\bar\theta\Gamma_M\zeta~,\ee where $\zeta$ is an
arbitrary local 32-component spinorial parameter and $\Gamma$ is as
in \cite{town88}:
$\Gamma=\f{1}{6\sqrt{-g}}\Gamma_{LMN}<E^L,E^M,E^N>$ satisfying
$\Gamma^2=1.$ In particular, taking the choice $\zeta =
(1+(\omega/\sqrt{-g})^{-1}(1+\Gamma)\kappa$ leads to a symmetry:
\be\delta_\kappa\theta=(1+\Gamma)\kappa, \quad \delta_\kappa X^M=
i\bar\theta\Gamma^M\delta_\kappa\theta, \quad \delta_\kappa\omega=
4i\f{\omega \sqrt{-g}}{\omega+\sqrt{-g}}(g^{-1})^{ij}
E^M_i\partial_j\bar\theta\Gamma_M\delta_\kappa\theta~,\ee where
$\kappa$ is an arbitrary local fermionic parameter so that the
transformations of $\theta$ and $X^M$ coincide with the
kappa-symmetry in \cite{town88};

c) Worldvolume diffeomorphisms: \be\delta_vX^M=v^i\partial_iX^M,
\quad \delta_v\theta=v^i\partial_i\theta,\quad
\delta_v\omega=\partial_i(\omega v^i)~,\ee where $v^i=\delta \xi^i$
is an arbitrary local bosonic parameter, and the Lagrangian
trans\-forms to a total derivative as $\delta_v{\mathcal
L}=\partial_i(v^i{\mathcal L})$.

Assume that the target space of the super-$p$-brane is a curved
supermanifold with $E^A_M(z)$ as its corresponding supervielbeins.
The $A=a,\alpha$ are the tangent space indices. Then the
super-$p$-brane action is given by \be\label{actbrback} S=-T_p\int
d^{p+1}\sigma(\sqrt{-\mbox{det}(E^a_iE^b_j\eta_{ab})}+\f{2}{(p+1)!}\varepsilon^{i_1\ldots
i_{p+1}}E^{A_1}_{i_1}\ldots E^{A_{p+1}}_{i_{p+1}}B_{A_{p+1}\ldots
A_1}),\ee where $E^A_i=\partial_iZ^M E^A_M$ is the pull-back of the
supervielbeins $E^A_M$. The field $B_{A_{p+1}\ldots A_1}(z)$ is the
superspace $p+1$-form potential. In fact, due to the
$\kappa$-symmetry of the action, only special values of $p$ and $D$
are allowable \cite{town88}. In this action the $p+1$-algebra also
can be introduced. Since \be
\mbox{det}(E^a_iE^b_j\eta_{ab})=\f{1}{(p+1)!}<E^{a_1},\ldots,
E^{a_{p+1}}><E_{a_1},\ldots, E_{a_{p+1}}>~,\ee
$$<E^{a_1},\ldots, E^{a_{p+1}}>=\varepsilon^{i_1\ldots i_{p+1}}E^{a_1}_{i_1}\ldots E^{a_{p+1}}_{i_{p+1}}~,$$
the action (\ref{actbrback}) can be reformulated in terms of the
Nambu $p+1$-brackets. The novelty of this reformulation is the
appearance of the Filippov--Nambu $p+1$-algebra.

As shown in \cite{park}, double dimensional reduction  of
supermembrane action (\ref{actNG}), putting $\xi^2=X^{10},
\Gamma^{(11)}=\Gamma^{10}~,$ straightforwardly leads  to the
well-known formulation of the type IIA superstring action by Green
and Schwarz \cite{gsw}. In a similar fashion to  type IIA
superstring action, the Schild version of type IIB superstring
covariant action in ten dimensions also appears. All the derivatives
therein appear through the Nambu brackets such that the two-algebra
structure of superstring theory and the three-algebra structure of
${\cM}$-theory become manifest. The Nambu two- and three-brackets
naturally arise since the dimensions of the string worldsheet and
the membrane worldvolume are two and three respectively \cite{park}.
One advantage to employ the Nambu brackets is the simplicity of the
double dimensional reduction: The three-bracket clearly reduces to
the two-bracket. Hence the Filippov--Lie $p+1$-algebra structure
becomes apparent for the super-$p$-brane theory. In  paper
\cite{park10} the authors have constructed supersymmetric extensions
of a bosonic $p$-brane action which reformulates the Nambu--Goto
action as an interacting multi-particle action with Filippov--Lie
$p$-algebra gauge symmetry.

The most intriguing question is: ''What is $\cM$-theory?'' It is
instructive to analyze the situation from the perspective of
spectrum of elementary excitations. Super\-strings describe massless
modes of lower spins $s \leq 2$ like graviton ($s$ = 2), gravitino
($s$ = 3/2), vector bosons ($s$ = 1) and matter fields with spins 1
and 1/2, as well as certain antisymmetric tensors. On the top of
that there is an infinite tower of massive excitations of all spins.
Since the corresponding massive parameter is supposed to be large,
massive higher spin excitations are not directly observed at low
energies. They are important however for the consistency of the
theory. Assuming that $\cM$-theory is some relativistic theory
admitting a covariant perturbative inter\-pre\-ta\-tion, we conclude
that it should necessarily contain higher spin modes to describe
superstring models as its particular vacua. There are two basic
alternatives: (i) $m \ne 0$: higher spin modes in $\cM$-theory are
massive or (ii) $m = 0$: higher spin modes in $\cM$-theory are
massless. Each of these alternatives is not straightforward. In the
massive case it is generally believed that no consistent superstring
theory exists beyond ten dimensions and therefore there is no good
guiding principle towards $\cM$-theory from that side. For the
massless option the situation is a sort of opposite: there is a very
good guiding principle but it looks like it might be too strong.
Indeed, massless fields of high spins are gauge fields. Therefore
this type of theories should be based on some higher spin gauge
symmetry principle with the symmetry generators corresponding to
various representations of the Lorentz group. It is very well known
however that it is a hard problem to build a nontrivial theory with
higher spin gauge symmetries. One argument is due to the
Coleman--Mandula theorem and its generalizations \cite{CM} which
claim that symmetries of S-matrix in a non-trivial (i.e.,
interacting) field theory in a flat space can only have sufficiently
low spins. These arguments convinced most of experts that no
consistent nontrivial higher spin gauge theory can exist at all.

However,  it was realized (see \cite{MVas} and references therein)
that the situation changes drastically once, instead of the flat
space, the problem is analyzed in the $AdS$ space with nonzero
curvature $\Lambda$. This generalization led to the solution of the
problem of consistent higher spin gravitational interactions in all
orders in interactions at the level of equations of motion. An
important general conclusion is that $\Lambda$ should necessarily be
nonzero in the phase with unbroken higher spin gauge symmetries
since it cancels the Coleman--Mandula argument which is hard to
implement in the $AdS$ background. However, up to date a fully
consistent action describing interactions of propagating higher spin
fields is not known. The nonlinear higher spin theory in four
dimensions was shown to be consistent up to cubic order at the
action level  and, later, at all orders at the level of equations of
motion. Concerning the problem of finding a consistent higher spin
action, it should be noted that one example does exist: the
Chern--Simons action in $d3$ constructed by Blencowe based on a
higher spin algebra \cite{blen} (see also \cite{sok}, \cite{sokHS}
in a related context) as the algebra of volume-preserving
diffeomorphisms $\partial_a(\sqrt{g}\Omega^{ab})=0$ of a manifold
$\cM$. This algebra is a subalgebra of the general diffeomorphisms
algebra of manifold $\cM$ and corresponds to the residual symmetry
of an extended object in the light-cone gauge. The symplectic
diffeomorphisms on $\cM^p$ are generated by
$L_\lambda=\Omega^{ab}\partial_b\lambda\partial_a,$ where $\lambda$
is an arbitrary function of $\sigma$ and
$\Omega^{ac}\Omega_{cb}=\delta^a_b.$ The generators $L_\lambda$ obey
the algebra $[L_{\lambda_1},L_{\lambda_2}]=L_{\lambda_3}$ where
$\lambda_{3}=\Omega^{ab}\partial_b\lambda_1\partial_a\lambda_2=\{\lambda_1,\lambda_2\}.$
Expanding the parameter $\lambda(\sigma)$ (whenever possible) in
terms of a complete infinite set of basis functions, one obtains an
infinite dimensional algebra from this composition law. This algebra
can be gauged by making the parameter $\lambda$ local,
$\lambda=\lambda(x,\sigma)$ in a spacetime, e.g. in a $2+1$. then
one can introduce a gauge field $\Gamma_m(x,\sigma)$ defined on
$\cM^3\otimes\cM^p$:
$\delta\Gamma_m=\partial_m\lambda+\{\Gamma_m,\lambda\}$. Next one
can write down an action for this field, in particular, a
Chern-Simons term \be S_{CS}=\int d^3x\int d^p\sigma
\sqrt{g}\varepsilon^{mnl}(\Gamma_m\partial_n\Gamma_l+\f13\{\Gamma_m,\Gamma_n\}\Gamma_l)~.\ee
It is important to realize that this action is invariant under the
gauge transformations only if the volume-preservation condition is
satisfied. This action was considered in \cite{sok} for the case of
a 2-sphere $S^2$ and 2-hyperboloid $H^2$. For example, for the case
of a 2-sphere it describes infinitely many spin-1 gauge fields,
while for a 2-hyperboloid it describes infinitely many higher spin
gauge fields including the gravitation field in $AdS$ space. In
order to reveal the infinite dimensional algebraic structure of the
algebra $SDiff(H^2)$, one needs an expansion on the 2-hyperboloid
$H^2$, such that the Lorentz transformation properties of the
generators will be manifest \cite{sokHS}. This can be done by using
a harmonic parametrization of $H^2$ \cite{gios} defined as follows.
Consider a set of variables $u^{\pm\alpha}$ parametrizing the group
$SL(2,R)\sim SO(2,1)$. The index $\alpha$ of $u^{\pm\alpha}$ is an
$SL(2,R)$ one, and the index $+$ or $-$ refers to a charge of the
$SO(2)$ subgroup of $SL(2,R)$ and $u^{+\alpha}u^-_\alpha=1$. The
coset $SL(2,R)/SO(2)$ will be realized on functions $f^{(q)}(u)$ of
$u^{\pm\alpha}$ having a define $SO(2)$ charge: $\partial^0
f^{(q)}=(u^{+\alpha}\partial_{u^{+\alpha}}-u^{-\alpha}\partial_{u^{-\alpha}})f^{(q)}=qf^{(q)}$.
In other worlds, those functions are given by the harmonic expansion
(for $q\geq 0$): \be f^{(q)}=\sum_{n=0}^\infty f^{(\alpha_1\ldots
\alpha_{n+q}\beta_1\ldots j_{n})}u^{+}_{\alpha_1}\ldots
u^+_{\alpha_{n+q}}u^-_{\beta_1}\ldots u^-_{\beta_n}~,\ee and
similarly for $q< 0$. Note that the coefficients in this expansion
are now irreducible spin-tensors of $SL(2,R)$. The two derivatives
$\partial/\partial \sigma^a$ on the coset are now represented by the
operators: \be\partial^{++}=u^{+\alpha}\f{\partial}{\partial
u^{-\alpha}}, \quad \partial^{--}=u^{-\alpha}\f{\partial}{\partial
u^{+\alpha}}, \quad [\partial^{++},\partial^{--}]=\partial^0~.\ee
Integration on $H^2$ is defined as follows: $\int du \cdot 1= 1,
\int du u^+_{(\alpha_1}\ldots u^+_{\alpha_{r}}u^-_{\beta_1}\ldots
u^-_{\beta_s)}=0$ for $r+s>0.$ This formal definition of the
integral has all the desired properties, in particular, it allows
integration by parts. Using the parametrization of $H^2$ introduced
above we can get an insight into the structure of the infinite
dimensional algebra of $SDiff(H^2).$ Now the composition law can be
rewritten in the following form
$\lambda_3=\partial^{++}\lambda_1(u)\partial^{--}\lambda_2(u)-(1\leftrightarrow
2).$ Expanding the harmonic functions we find
$$\lambda_{12}=\sum_{r,s,k}C_k^{2r,2s}\lambda_1^{\alpha_1\ldots \alpha_k\beta_{k+1}\ldots \beta_{2r}}
\lambda^{\alpha_{k+1}\ldots\alpha_{2s}}_{2\ \ \ \ \ \ \
\beta_{k+1}\ldots\beta_{2r} }u^+_{(\alpha_1}\ldots
u^+_{\alpha_k}u^-_{\alpha_{k+1}}\ldots
u^-_{\alpha_{2s)}}-(1\leftrightarrow 2),$$ where $C_k^{2r,2s}$ are
$SL(2,R)$ Clebsh--Gordon coefficients. Note that this composition
law for area-preserving diffeomorphisms of $H^2$ had be already
given in \cite{sok}. In order to make contact with familiar
Chern--Simons theories based on finite dimensional spacetime
algebras, one must examine the finite dimensional truncation of
$SDiff(H^2).$ Now the generalization the concepts discussed above to
the case of a supermanifold is straightforward  \cite{sokHS}. If we
take $\Gamma_m=(\omega^{\alpha\beta}_m +\theta^4
e^{\alpha\beta}_m)u^+_{(\alpha}u^-_{\beta)}~,$ the Chern--Simons
action is just the action for the Poincare gravity in $2+1$
dimensions \be S=\int
d^3x\varepsilon^{mnl}(e^{\alpha\beta}_m\partial_n\omega_{l\alpha\beta}-
e^\alpha_{m\beta}\omega^\beta_{n\rho}\omega^\rho_{l\alpha})~.\ee
Since the Chern-Simons theory is a true gauge theory, the resulting
higher spin theory (how shown in the above toy example $H^2$) is
consistent by construction and naturally extends the
Einstein-Hilbert action (which in $d$3 also has an interpretation as
a Chern-Simons action). It is, however, only of limited use since it
is topological theory that does not give rise to propagating degrees
of freedom. It is possible to describe a super $AdS$ (also
superconformal) field theory for an infinite tower of integer and
half-integer higher spin field in a similar fashion.

As a review of the construction of consistent higher-spin
four-dimensional theories based on 5$d$ topological theory with
Chern-Simons actions, see \cite{hohm} and the references therein.

\section{Conclusions and outlook}
A (1+2)-dimensional relativistic gauge theory based on the Filippov
3-algebra \cite{fil}  rather than on the Lie algebra, was proposed
recently by BLG \cite{BLG}, as a model of multiple M2-branes. The
model has an $OSp(8|4)$ conformal symmetry  as expected for the
infra-red fixed point of the Yang--Mills type gauge theory on
coincident D2-branes. The construction requires a metric on the
3-algebra and if this metric is positive definite then the structure
constants of the 3-algebra define a totally-antisymmetric
fourth-rank tensor satisfying a fundamental identity. When the
struc\-ture constants vanish one has a 'trivial' 3-algebra and the
model reduces to a free theory for the $\cN = 8$ scalar multiplet,
as expected for the conformal limit of a single planar M2-brane. A
non-trivial realization based on the Lie algebra $so(4)$ was given
by BLG, and it appears to describe two coincident M2-branes on an
orbifold $\mathbb{R}^8/\mathbb{Z}_2$. Other possibilities emerge
when one allows for Lorentzian metrics on the 3-algebra \cite{Lor},
\cite{quantLor} but these models have ghosts. Various other facets
of  BLG models have been addressed in other papers; an incomplete
list can be found in \cite{Gai}-\cite{schnabl}. In the context of
the original BLG model, with positive definite metric, there remains
one other possibility: there is an infinite-dimensional realization
of the 3-algebra in terms of the Nambu bracket on a
three-dimensional space. In this realization, the BLG model is
essentially an exotic gauge theory for the group of
volume-preserving diffeomorphisms of this space, where by 'exotic'
we mean that the gauge theory is not of the Yang--Mills type. This
is not the first occasion on which exotic gauge theories based on
volume preserving diffeomorphisms have appeared. They also arise
from light-cone gauge fixing of relativistic $p$-brane actions for
$p > 2$; these are exotic gauge theories with a group of
$p$-volume-preserving diffeomorphisms, $SDiff_p$, as the gauge group
\cite{bandos}. As is well-known, the flux of the 2-form potential on
the M5-brane may be interpreted as M2-branes 'dissolved' in the
M5-brane. Thus, a single M5-brane may contain multiple M2-branes and
is therefore a promising starting point for a construction of the
BLG model for multiple M2-branes. Another indication of this is that
the Nambu-bracket realization of the BLG theory introduces some
'internal' Riemannian 3-manifold $M_3$, so that the 'total' space
dimension is $2 + 3 = 5.$ In fact, it has been proposed in recent
papers that the Nambu-bracket realization of the BLG model is
equivalent to the M5-brane action \cite{ho}. One of the many obvious
question is whether analogous results might emerge by considering
M5-branes of other topologies, for example $S^1\times M_4$ for some
closed 4-manifold $M_4.$ One might imagine that this could be
related to some exotic $(1+1)$-dimensional gauge theory based on the
Filippov 4-algebra. A natural question is whether there exist gauge
theories for which the gauge group is the group $SDiff(M_n)$ of
volume-preserving diffeomorphisms of some $n$-dimensional manifold
$M_n$ for $n \geqslant 3$; we assume that $M_n$ is closed and
compact with respect to some volume $n$-form. Another outstanding
problem is the nature of the 6$d$ conformal field theory governing
the low energy dynamics of $N$ coincident M5-branes. In light of
what we now know about multiple coincident M2-branes, it seems
likely that this problem will simplify in the $N \rightarrow\infty$
limit. Given that a condensate of M2-branes may be viewed, in some
sense, as an M5-brane, then is there a similar sense in which an M5
condensate could be viewed as a yet higher-dimensional M-brane?
Recalling that the recent advances in the M2 case were prompted by
the Basu--Harvey proposal that the boundary of multiple M2-branes on
an M5-brane might be understood in terms of fuzzy 3-spheres, it is
natural to reconsider the implications of the recent demonstration
\cite{open5} that an M5-brane can have a boundary on an M9-brane,
which is a boundary of the 11-dimensional bulk spacetime of
$\cM$-theory. In this context we should mention that
higher-dimensional generalizations of the Basu--Harvey equation have
been considered in \cite{zabz}. In general, could say that
$\cM$-theory  must be a new kind of theory, which should perhaps be
formulated in terms of completely new degrees of freedom, and
requires new physical principles.  Now it seems the main aim of this
development programme is to put the Nambu-bracket realization of the
BLG theory into a larger context by developing further the general
principles of $SDiff$ gauge theory.

Though this a brief survey of some important trends in recent
investigations poses more questions that provides the answers, we
feel that the subject of the Filippov--Nambu higher algebraic
operations might be relevant for future development of mathematical
structure related to a great many physical problems and then
certainly deserve further studies.

Author is very grateful to J.A. de Azcarraga, O. Hohm, C. Krishnan,
H. Lin, J.-H. Park and C. Zachos    for valuable comments.

%\section{References}
\bigskip

\end{document}